\documentclass[a4paper,fleqn,usenatbib,useAMS]{mnras}

\usepackage{graphicx}	
\usepackage{amsmath}	
\usepackage{amssymb}	
\usepackage{multicol}        
\usepackage{bm}		
\usepackage{pdflscape}	
\usepackage{footnote}
\usepackage{color}

\newcommand{\be}{\begin{equation}}
\newcommand{\ee}{\end{equation}}
\newcommand{\ba}{\begin{eqnarray}}
\newcommand{\ea}{\end{eqnarray}}

\title[Testing photo-z measurements with CSS-OS filter definition]{Testing photometric redshift measurements with filter definition of the Chinese Space Station Optical Survey (CSS-OS)} 

\author[Ye Cao et al.]{Ye Cao$^{1,2}$, Yan Gong$^{1}$ \thanks{E-mail: gongyan@bao.ac.cn}, Xian-Min Meng$^3$, Cong K. Xu$^{4,5}$, Xuelei Chen$^{1,2,6}$, 
\newauthor Qi Guo$^1$, Ran Li$^{1,2}$, Dezi Liu$^7$, Yongquan Xue$^{8,9}$, Li Cao$^3$, Xiyang Fu$^3$, 
\newauthor Xin Zhang$^3$, Shen Wang$^3$, Hu Zhan$^3$
 \\
$^{1}$ Key Laboratory for Computational Astrophysics, National Astronomical Observatories, Chinese Academy of Sciences, 20A Datun \\Road, Beijing 100012, China\\
$^{2}$ School of Astronomy and Space Sciences, University of Chinese Academy of Sciences, Beijing 100049, China\\
$^{3}$ Key Laboratory of Space Astronomy and Technology, National Astronomical Observatories, Chinese Academy of Sciences, Beijing\\ 100012, China\\
$^{4}$ National Astronomical Observatories, Chinese Academy of Sciences, 20A Datun Road, Beijing 100012, China\\
$^{5}$ South American Center for Astronomy, CAS, Camino El Observatorio 1515, Las Condes, Santiago, Chile\\
$^{6}$ Center for High Energy Physics, Peking University, Beijing 100871, China\\
$^{7}$ Department of Astronomy, School of Physics, Peking University, Beijing 100871, China\\
$^{8}$ CAS Key Laboratory for Research in Galaxies and Cosmology, Department of Astronomy, University of Science and Technology of\\ China, Hefei 230026, China\\
$^{9}$ School of Astronomy and Space Science, University of Science and Technology of China, Hefei 230026, China\\}

\date{Accepted XXX. Received YYY; in original form ZZZ}

\pubyear{2018}

\begin{document}
\label{firstpage}
\pagerange{\pageref{firstpage}--\pageref{lastpage}}
\maketitle

\begin{abstract}
The Chinese Space Station Optical Survey (CSS-OS) is a major science project of the Space Application System of the China Manned Space Program. This survey is planned to perform both photometric imaging and slitless spectroscopic observations, and it will focus on different cosmological and astronomical goals. Most of these goals are tightly dependent on the accuracy of photometric redshift (photo-$z$) measurement, especially for the weak gravitational lensing survey as a main science driver. In this work, we assess if the current filter definition can provide accurate photo-$z$ measurement to meet the science requirement. We use the COSMOS galaxy catalog to create a mock catalog for the CSS-OS. We compare different photo-$z$ codes and fitting methods that using the spectral energy distribution (SED) template-fitting technique, and choose to use a modified LePhare code in photo-$z$ fitting process. Then we investigate the CSS-OS photo-$z$ accuracy in certain ranges of filter parameters, such as band position, width, and slope. We find that the current CSS-OS filter definition can achieve reasonably good photo-z results with $\sigma_z\sim0.02$ and outlier fraction $\sim$3\%.
\end{abstract}

\begin{keywords}
cosmology: observations - theory - large-scale structure of universe
\end{keywords}

\section{Introduction}
Photometric sky survey is a basic observation for modern astronomy research. The positions of large number of targets on the celestial sphere and their fluxes in several wavelength bands defined by photometric filters are obtained by such surveys. A number of ongoing and planned large photometric surveys with large areas and deep fields are well known, e.g. the Sloan Digital Sky Survey (SDSS)\footnote{\tt http://www.sdss.org/} \citep{Fukugita96,York00}, the Large Synoptic Survey Telescope (LSST) \citep{Ivezic08,LSST09}, the Euclid space telescope \citep{Laureijs11}, Dark Energy Survey (DES)\footnote{\tt https://www.darkenergysurvey.org/}, Javalambre Physics of the Accelerating Universe Astrophysical Survey (J-PAS)\footnote{\tt http://www.j-pas.org/}\citep{Benitez14}, etc. These surveys can provide us much information on the spatial distribution, clustering, as well as gravitational lensing of galaxies, which are very valuable in solving a number of fundamental problems, such as the properties of dark energy and dark matter, the origin of the Universe, and formation and evolution of galaxies \citep{Lenz98,Richards02,Budavari03,Helmi03,Ross12}. 

To achieve all of these scientific goals, the redshift information of galaxies is needed. However, high precision measurements of the galaxy redshift require spectroscopic redshift measurements, which is quite time-consuming, especially for the cosmological studies with large sample of galaxies. Nevertheless, an estimate of the redshift can be obtained from the photometric survey, that is much more efficient than spectroscopic survey given the same survey area and depth. Although not as precise as the spectroscopic redshift, in a number of important applications, e.g. weak gravitational lensing, the photometric redshift (photo-$z$) is adequate for current studies. Furthermore, it is also very useful when selecting a sub-sample of targets for a spectroscopic survey. Thus, the photometric survey is very useful in the current cosmological studies. For a photometric survey, it is important to study if a filter set can provide accurate photo-$z$ measurement to satisfy its science requirement. Besides, understanding how the photo-$z$ accuracy varies as parameters of photometric filter changing is also quite helpful for designing a good set of photometric filter system to improve photo-$z$ calibration.

In this work, we investigate the photo-$z$ measurements with filter definition of the imaging part of the Chinese Space Station Optical Survey (CSS-OS). As a major science project established by the Space Application System of the China Manned Space Program, this survey will be performed by a 2-meter telescope operating in the same orbit of the China Manned Space Station. It includes both photometric imaging and slitless spectroscopic observations. It shall have a large field of view $\sim1$ deg$^2$, high spatial resolution $\sim0.15$ arcsec, faint magnitude limits, and wide wavelength coverage from near-ultraviolet (NUV) to near infrared (NIR) bands \citep{Zhan11}. It would be the basis for many kinds of cosmological and astronomical observations, including weak gravitational lensing, baryon acoustic oscillation, galaxy and galaxy clusters, active galactic nuclei, etc. Fundamental questions about gravity, dark mater and dark energy, the cosmic large scale structure, galaxy formation and evolution, the formation of super-massive black hole and so on could be investigated. 

Most of these scientific goals are heavily dependent on the accuracy of photometric redshift, especially for the weak gravitational lensing survey as a main science driver of the CSS-OS. According to previous studies, the photo-$z$ accuracy for future photometric weak lensing surveys needs to achieve $\sigma_z<0.05$ at least and $\sigma_z\simeq 0.02$ as a goal \cite[e.g. see][]{LSST09,Zhan06}. Hence, it is necessary to explore if the current filter definition of the CSS-OS (in certain parameter ranges) can offer accurate photo-$z$ estimate to meet the science requirement. This study can also provide a guidance of the filter design for other similar surveys.

Many methods of estimating redshift from the photometric data have been developed over the years. Roughly speaking,  they can be classified as two types. One type may be called ``template fitting'' method \citep{Lanzetta96,Fernandez99}, which extracts redshift by fitting photometric data with the templates of galaxy spectral energy distributions (SEDs). Publicly available codes of this type include Hyperz \citep{Bolzonella00}, BPZ \citep{Benitez00}, ZEBRA \citep{Feldmann06}, EAZY \citep{Brammer08}, LePhare \citep{Arnouts99,Ilbert06}, etc. Another type may be called ``training set'' method \citep{Connolly95,Brunner97}, which obtains an empirical relation between redshift and galaxy properties (e.g. magnitude and color) using a galaxy sample with measured spectroscopic redshifts. The neutral network code ANNz \citep{Firth03,Collister04} is of this type. These two approaches have different advantages. In this work, we adopt the SED fitting technique, since the magnitude limit of the CSS-OS is much higher than usual spectroscopic surveys, that it is hard to find suitable training set can be used to perform this study.

In order to simulate the observational data as real as possible, we make use of the COSMOS galaxy catalog \citep{Ilbert09}, whose magnitude limit is similar to the nominal value of the CSS-OS, and therefore also has similar galaxy redshift distribution, magnitude distribution, and galaxy types, though the CSS-OS would cover much wider sky area. Using this catalog, we select sub-sample based on the CSS-OS instrumental parameters with high data quality, generate mock flux data for each filter passband, and estimate the observational errors. After comparing different photo-$z$ fitting codes, we choose to use a modified LePhare code to perform the filter calibration with two fitting methods. We find they can substantially suppress the fraction of catastrophic redshift. Then we discuss the effect of each band on photo-$z$ accuracy by omitting them, and investigate the CSS-OS photo-$z$ accuracy in three parameter ranges of filter transmission curve, i.e. the position of central wavelength, the band wavelength width, and the slope of transmission curve. We also compare the CSS-OS filters with other types of filters using real CSS-OS detector efficiency.

This paper is organized as follows: in Section 2, we introduce the filter definition of CSS-OS and the method used to create the mock data. In Section 3, we explore three currently widely-used photo-$z$ fitting codes, i.e. LePhare, EAZY and Hyperz, and modify the LePhare code to include the information of poorly detected data. In Section 4, we investigate the dependency of photo-$z$ accuracy on each CSS-OS filter and the three filter transmission parameters, and compare the results with other filter sets. We finally summarize the results in Section 5.

\section{Mock galaxy flux data}
In this section, we first introduce the CSS-OS filter definition, then discuss the galaxy catalog we use to mimic the CSS-OS observations, and finally estimate the mock flux and error from galaxy SED models.

\subsection{The CSS-OS filter definition}

\begin{figure*}
\centering
\includegraphics[width=1.0\columnwidth]{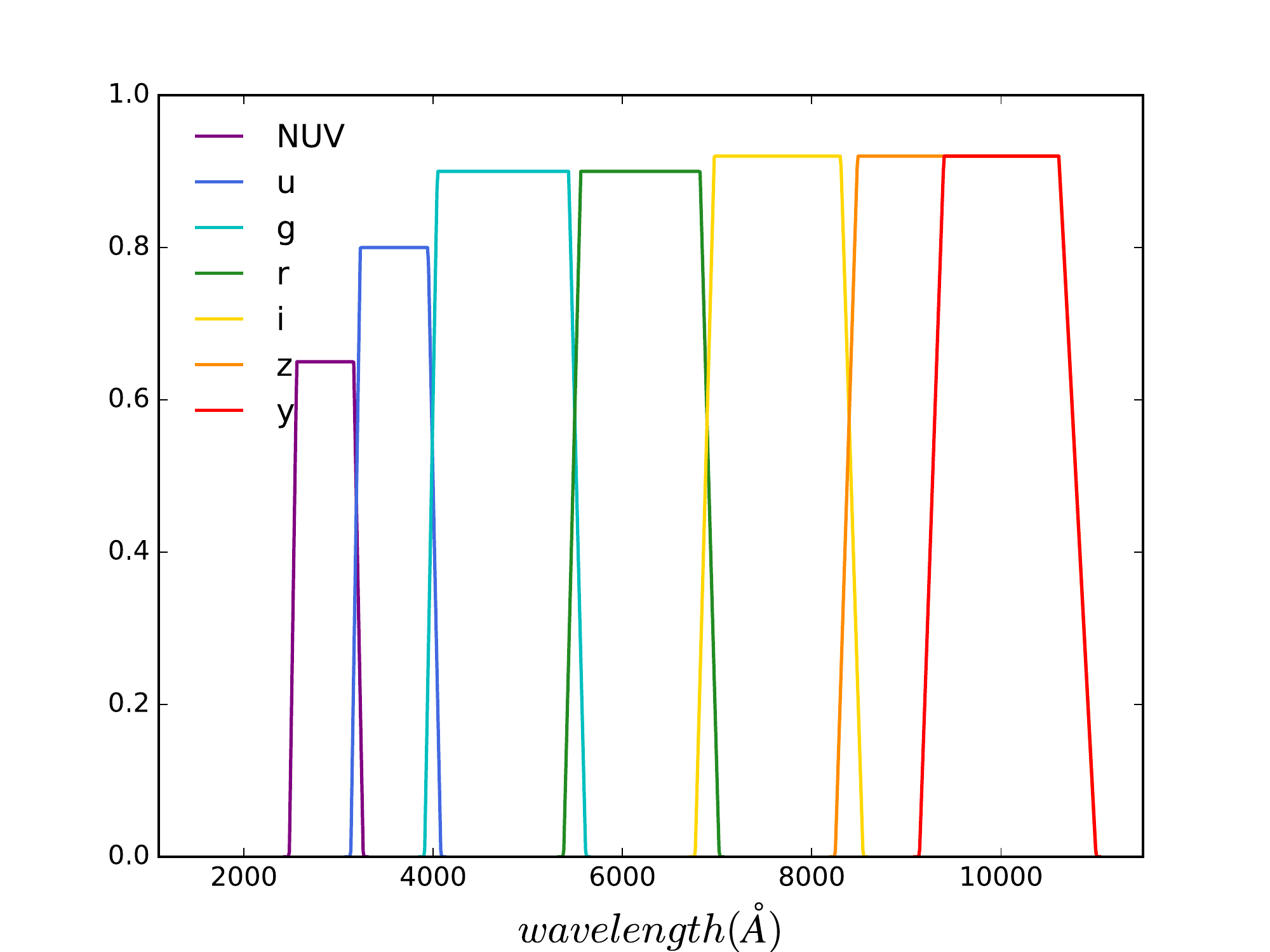}
\includegraphics[width=1.0\columnwidth]{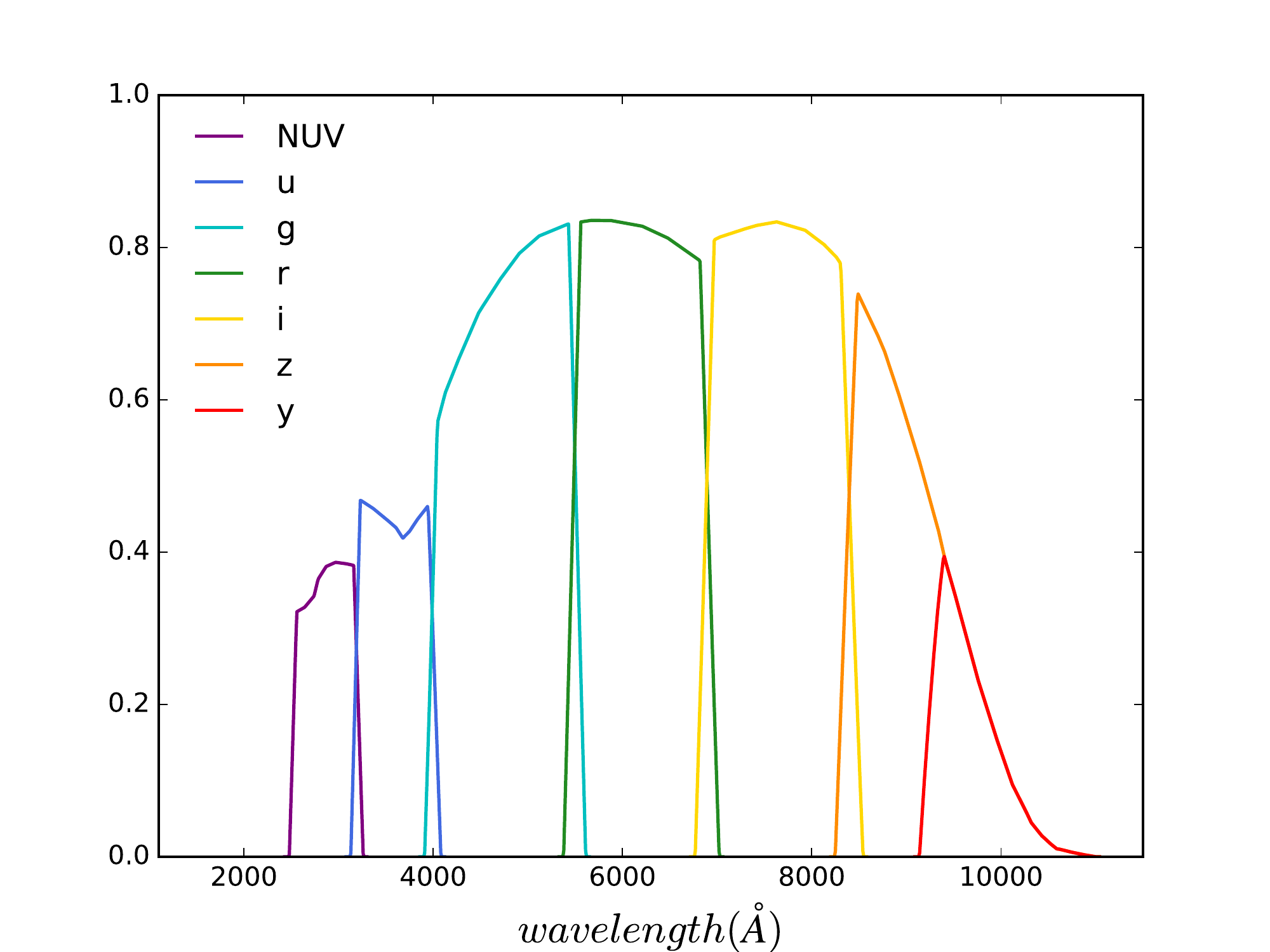}
\caption{$Left:$ The intrinsic transmission curves for the seven filters from NUV to NIR bands, including the $NUV$, $u$, $g$, $r$, $i$, $z$, and $y$ bands. $Right:$ The total transmission by considering detector quantum efficiency.}
\label{fig:filters} 
\end{figure*}

Based on current telescope instrumental design, the CSS-OS will be conducted with a combination of 7 broadband filters, and 
their nominal AB magnitude 5$\sigma$ limits are $NUV\simeq$25.4, $u\simeq$25.5, $g\simeq$26.2, $r\simeq$26.0, $i\simeq$25.8, 
$z\simeq25.7$ and $y\simeq25.5$ for point sources (measured within 80\% energy concentration region of the CSS-OS Gaussian-shape PSF). In Figure {\ref{fig:filters}}, we show the transmission curves for seven filters that are under test. The left panel shows the intrinsic transmission curves, and the right panel gives the total transmission including the detector quantum efficiency. The definition parameters for the filters are listed in Table \ref{tab:filters}. Here we show the mean wavelength $\lambda_{\rm mean}$, full width at half maximum (FWHM), and the wavelengths at 1\% and 90\% of the maximum transmission curve (i.e. left: $\lambda_{\rm -01}$ and $\lambda_{\rm -90}$, and right: $\lambda_{\rm +90}$ and $\lambda_{\rm +01}$). The top transmission efficiency for each band is estimated to be 65\% for $NUV$, 80\% for $u$, 90\% for $g$ and $r$ bands, and $92\%$ for $i$, $z$ and $y$ bands, respectively. The current CSS-OS filters are designed to be similar with the corresponding filters of the SDSS and LSST, especially for the $u$, $g$, $r$, and $i$ bands, which are well developed and tested by practice. This can be helpful for the CSS-OS to achieve its scientific goals, and easy to compare its observational results, e.g. magnitudes and colors of observed objects, with other surveys. This filter definition is the basic case which is used in our following discussion, and we will study the accuracy of photo-$z$ it can reach with the galaxy catalog we adopt as described below.

\begin{table*}
\centering
\caption{The CSS-OS filter definition.}
\label{tab:filters}
\begin{tabular}{ c  c  c  c  c  c  c  c}
\hline\hline
Filter & $\lambda_{\rm mean}$ $(\AA)$ & FWHM $(\AA)$ & $\lambda_{\rm -01}$ $(\AA)$ & $\lambda_{\rm -90}$ $(\AA)$ & $\lambda_{\rm +90}$ $(\AA)$ & $\lambda_{\rm +01}$ $(\AA)$ & Trans.\\
\hline
$NUV$ &  2877 &701 & 2480 & 2550 & 3170 & 3260 & 65\%\\
$u$ &  3595 & 847& 3130 & 3220 & 3960 & 4080 & 80\%\\
$g$ & 4798 & 1562 & 3910 & 4030 & 5450 & 5610 & 90\%\\
$r$ & 6186 & 1471& 5380 & 5540 & 6840 & 7020 & 90\%\\
$i$ & 7642 & 1577& 6770 & 6950 & 8330 & 8540 & 92\%\\
$z$ & 9046 & 2477 & 8250 & 8460 & 10650 & 11000 & 92\%\\
$y$ & 9654 & 1576 & 9140 & 9370 & 10650 & 11000 & 92\%\\
\hline
\end{tabular}
\end{table*}

\subsection{Galaxy Catalog}

\begin{table}
\centering
\caption{The fliter, telescope, effective wavelength, and width used in our COSMOS catalog.}
\label{tab:catalog_filter}
\begin{tabular}{ c  c  c  c  }
\hline\hline
Filter & telescope & effective $\lambda$ & FWHM $(\AA)$  \\
\hline
$u*$ &  CFHT &3911.0 & 538.0  \\
$B_J$ &  Subaru & 4439.6& 806.7  \\
$V_J$ & Subaru & 5448.9 & 934.8  \\
$g^+$ & Subaru & 4728.3& 1162.9  \\
$r^+$ & Subaru & 6231.8& 1348.8  \\
$i^+$ & Subaru & 7629.1 & 1489.4  \\
$z^+$ & Subaru & 9021.6 & 955.3  \\
$i*$ & CFHT & 7628.9& 1460.0  \\
$J$ & UKIRT & 12444.1 & 1558.0  \\
$K$ & CFHT & 21480.2 & 3250.0  \\
\hline
\end{tabular}
\end{table}

\begin{figure}
\centering
\includegraphics[width=1.1\columnwidth]{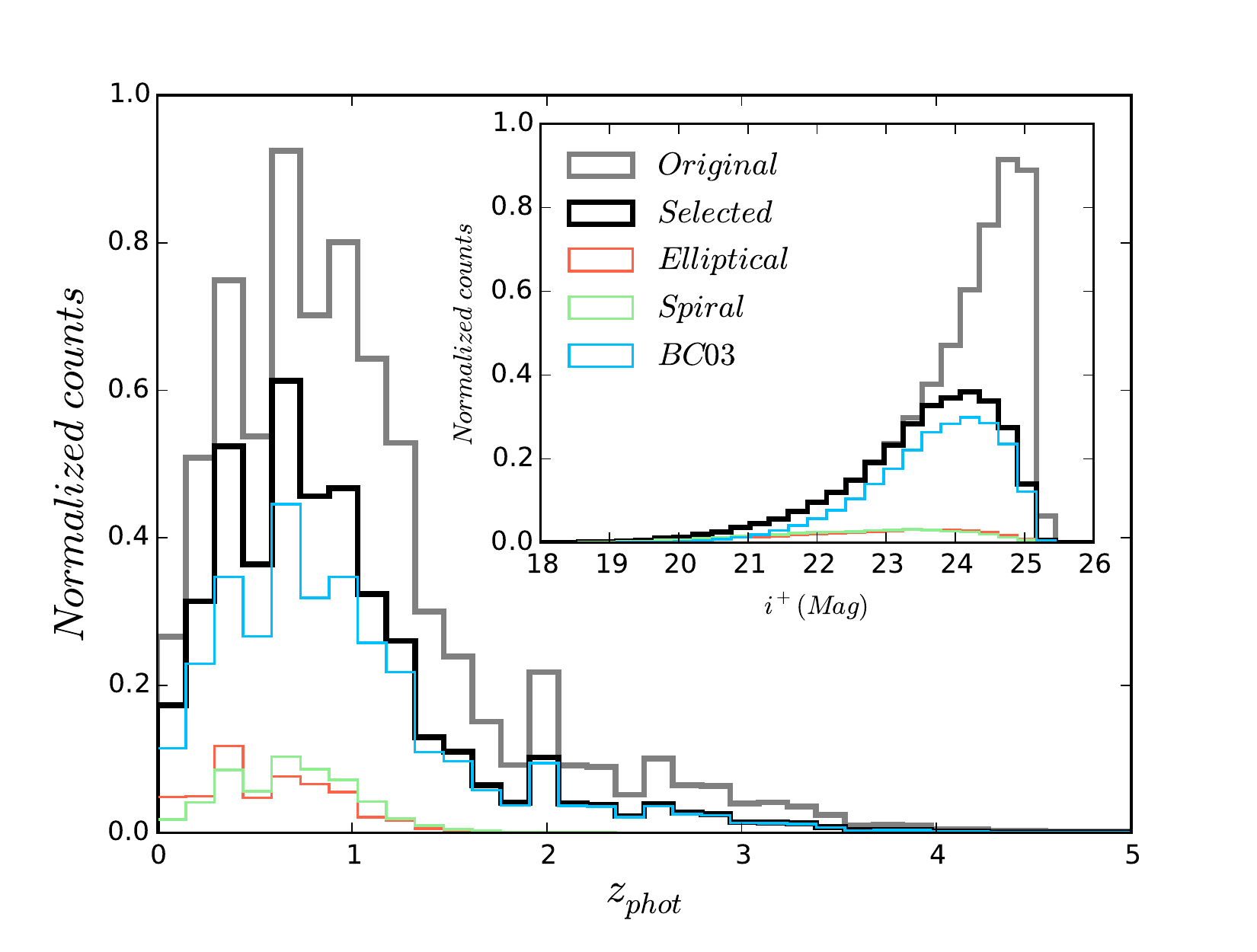}
\caption{The redshift and magnitude distributions of the COSMOS galaxy catalog we use \citep{Ilbert09}. The gray and black histograms show the original and high-quality selected galaxy samples, respectively. The distributions of elliptical, spiral, and young blue star forming (fitted by the BC03 method, and labeled as BC03 following \citet{Ilbert09}) galaxies of the selected sample are also shown in red, green and blue histogram, respectively.}
\label{fig:zm_dis}
\end{figure}

In order to study the photo-$z$ calibration for the CSS-OS, we first need to find a galaxy catalog which can represent the survey. It should have similar redshift and magnitude distributions as expected as the CSS-OS. Since the magnitude limit of the CSS-OS can achieve $i\sim26$ for point source with 5$\sigma$ detection, it is expected to be about one magnitude brighter for galaxy surface sources. Here we make use of the COSMOS galaxy catalog \citep{Capak07,Ilbert09} from an accurate photo-$z$ survey in 2-deg$^2$ COSMOS field, covering the near-UV, optical, and near-IR bands. This catalog contains about 380,000 sources with $i^+ \le 25.2$, which is obtained by Subaru Telescope \citep{Taniguchi07,Capak08,Taniguchi09}, Canada-France-Hawaii Telescope (CFHT) \citep{Boulade03}, and United Kingdom Infrared Telescope (UKIRT). The details of this catalog, such as band coverage and origin, can be found in Table~\ref{tab:catalog_filter}.
After removing stars, X-ray, and masked sources, we obtain about 219,000 galaxies as our original galaxy catalog.  This original catalog includes necessary information for generating mock data, such as the redshift, magnitude, galaxy type, and galaxy size. In order to perform the photo-$z$ fitting with required accuracy, we also need to select sources with high data quality. Here we calculate the signal to noise ratio (SNR) for each galaxy in the catalog based on the CSS-OS instrumental parameters (see Section~2.3 for details of the SNR estimate), and select the sources with SNR$\ge$10 in $g$ or $i$ band. Then we get $\sim$126,000 sources ($\sim$58\% of the original galaxy catalog) after the selection\footnote{For the CSS-OS, we find that about 208,000 galaxies ($\sim95\%$ of the original sample) can be selected if requiring a typical value of the photo-$z$ fitting variance $\sigma_z\lesssim0.05$ (defined in Section 3) in weak lensing survey.}. Finally, 10,000 galaxies are randomly selected from the high-quality sample, which has similar redshift and magnitude distributions, and they will be used in our photo-$z$ calibration. 

The redshift and magnitude distributions of the catalogs are shown in Figure \ref{fig:zm_dis}. We show the original and selected samples in gray and black histograms, respectively. We can see that the redshift distribution of the selected sample has a peak around $z=0.7$, and can extend to $z\sim4$. The peak of its magnitude distribution is at $i\sim24.2$, and the range is from $i\sim19$ to 25. As compared to the distributions of the original sample, we find that the majority of faint sources with $i\gtrsim24$ is removed in the selected sample. 

The redshift and magnitude distributions of elliptical, spiral, and young blue star forming (fitted by the method given in \citet{Bruzual03}, hereafter BC03) galaxies are also shown in red, green and blue histograms, respectively. We find that the percentages of these three kinds of galaxies are 75\% for young blue star forming, 13\% for spiral, and 12\% for elliptical galaxies, which means the young blue star forming galaxies are dominant in the selected catalog. The peak of redshift distributions for both young blue star forming and spiral galaxies is at $z\sim0.7$, which is the same as the total distribution, while it is around $z=0.3$ for elliptical galaxies.

\subsection{Flux and error estimation}

For a given galaxy in the catalog, mocked flux in each band can be calculated by convolving the galaxy redshifted SED with the filter response or transmission function, which is given by
\be \label{eq:f_obs}
F^{\rm mock}_x = \int S_{\rm model}(\lambda) T_x(\lambda) d\lambda.
\ee
Here $T_x$ is the response function for band $x$, $S_{\rm model}(\lambda)$ is the galaxy SED derived from SED model (based on the LePhare template that fits best the COSMOS data, with extrapolations and modifications that will be discussed in the following), and $\lambda=\lambda_{\rm res}(1+z)$ where $\lambda_{\rm res}$ is the rest-frame wavelength. Then $F^{\rm mock}_x$ is rescaled according to the $i$ band apparent magnitude given by the COSMOS galaxy catalog. Note that there is difference between the $i$ bands used in the COSMOS catalog (Subaru $i^+$ is used) and in our survey. We have converted the $i^+$ band flux from the COSMOS catalog to the CSS-OS $i$ band flux by $F^{\rm obs}_{i}=F^{\rm obs}_{i^+}F^{\rm mock}_{i}/F^{\rm mock}_{i^+}$, where $F^{\rm obs}$ is the observational flux, and $F^{\rm mock}$ is the mock flux calculated by Eq.~(\ref{eq:f_obs}).

\begin{figure}
\includegraphics[width=1.05\columnwidth]{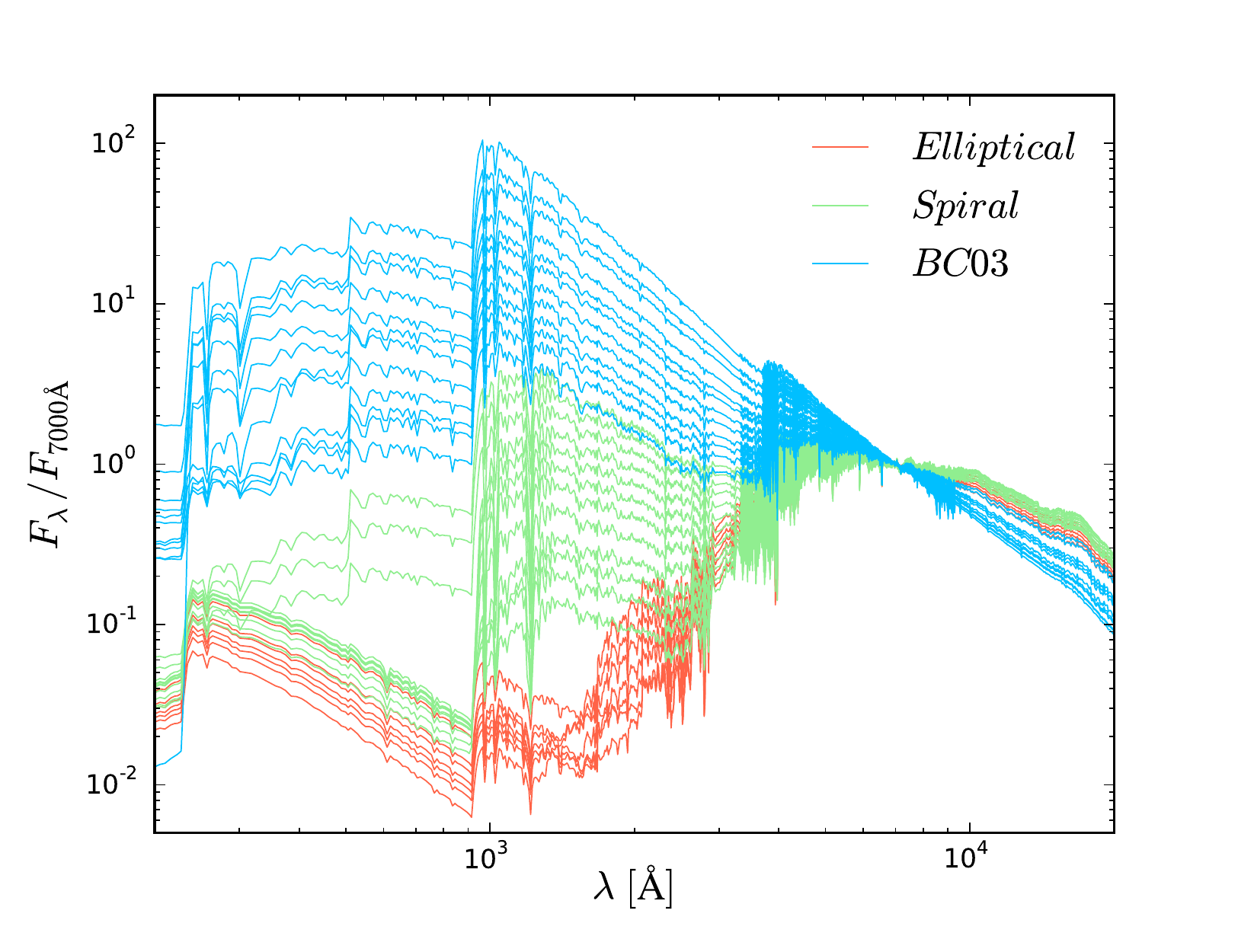}
\caption{The SED templates we use in the analysis. The SEDs of young blue star forming, spiral, and elliptical galaxies, which are derived from the BC03 method and the templates given in \citet{Polletta07}, are shown in blue, green and red curves, respectively. These templates are taken from the COSMOS template library \citep{Arnouts99,Ilbert06}, and we extend them from $\sim900$ $\rm\AA$ to $\sim90$ $\rm\AA$ using the BC03 method.}
\label{fig:SED} 
\end{figure}

In Figure \ref{fig:SED}, we show the intrinsic SED templates that we used to generate mock flux data. They are based on templates taken from the COSMOS template library \citep{Arnouts99,Ilbert06}, which include seven templates for elliptical galaxy, twelve for spiral galaxies (from S0 to Sdm), and twelve for young blue star forming galaxy (starburst ages from 0.03 to 3 Gyr) \citep{Ilbert09}. The templates of elliptical and spiral galaxies are derived from the templates in \citet{Polletta07}, and the young blue star forming galaxy templates are generated by the BC03 models. 

Since the CSS-OS covers large redshift and wavelength ranges, we extend the wavelength coverage of these templates from $\sim900$ $\rm\AA$ to $\sim90$ $\rm\AA$ using the BC03 method. We fit each SED template with 45 BC03 simple stellar population (SSP) templates in the wavelength range of 915--15000\,$\rm\AA$ to obtain the stellar components of the SED templates, and select 15 ages from 1\,Myr to 13\,Gyr of BC03 SSPs. The subsolar, solar, and supersolar metallicities of the 15 ages are adopted, and we take $Z=0.004$, $0.02$, and $0.05$ for the three cases, respectively. The spectral fitting gives excellent result with reduced $\chi^2 = 0.1\sim1.5$. The SED templates can be rebuilt based on the fractions of stellar components and other relevant factors. Finally, the extension of SED templates is achieved by extracting the spectra from the rebuilt SEDs in the wavelength range of 90\,$\rm\AA$--160\,$\mu \rm m$.

For each galaxy, the COSMOS catalog provides the corresponding best-fit SED template shown in Figure~\ref{fig:SED}, and then we can conveniently generate our mock flux data based on it. However, note that here we do not directly adopt the SED templates when generating the mock data. In order to avoid ``over-fitting" effect in the photo-z fitting process\footnote{The over-fitting effect appears when using the same set of SED template to fit the mock flux data generated from it.}, we generate new SED for each galaxy by linearly combing the SED templates. A random Gaussian factor $a$ (centered at 0 with $\sigma=0.1$) is used for linear combination of the SED template with its adjacent templates,
\be
S_{\rm int} = (1-a) S_{\rm i} + a\,S_{\rm i\pm1},
\ee
where $S_{\rm int}$ is the intrinsic galaxy SED used in the mock data, $S_{\rm i}$ is the best-fit SED template for a galaxy given by the COSMOS catalog, and $S_{\rm i\pm1}$ is the adjacent templates of $S_{\rm i}$. We take $S_{\rm i+1}$ for $a>0$, and $S_{\rm i-1}$ when $a<0$.

Note that, in real survey, the SED templates we take are probably incomplete, since a large fraction of the galaxies in the CSS-OS sample would be too faint to have been thoroughly studied before. Thus, an extra uncertainty of photo-$z$ fitting will be introduced due to this SED ``incompleteness problem". The detailed study of this problem is beyond the scope of this work, and a simple test about this uncertainty can be found in \cite{Abrahamse11}.

\begin{figure}
\includegraphics[width=1.05\columnwidth]{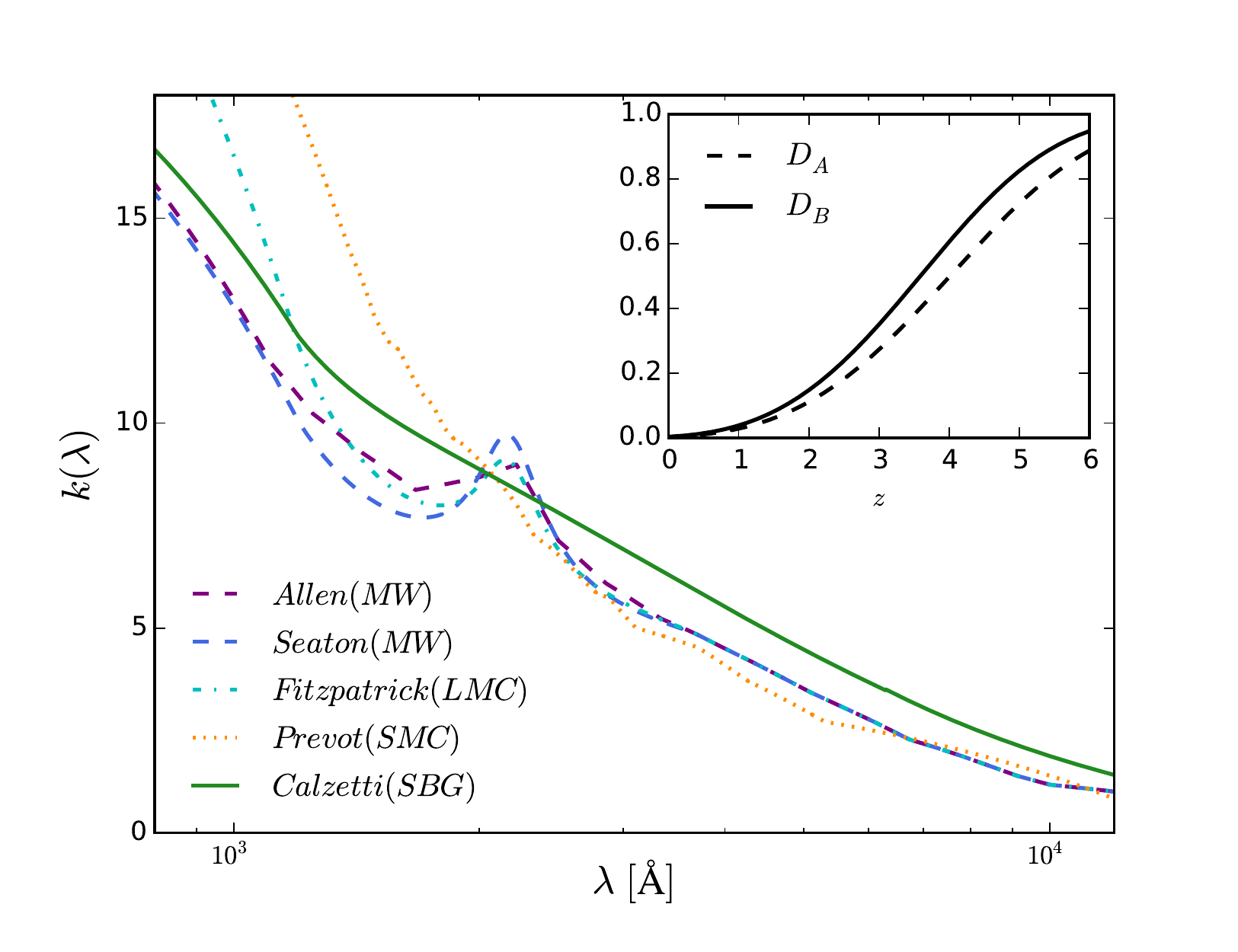}
\caption{The dust extinction laws we adopted in this work. Five extinction laws are shown here, which are obtained from the studies of the Milky Way (MW), Large Magellanic Cloud (LMC), Small Magellanic Cloud (SMC), and starburst galaxy (SBG)\citep{Allen76,Seaton79,Fitzpatrick86,Prevot84,Bouchet85,Calzetti00}. The absorption by the IGM is also considered, and we use the average flux decrement factors $D_{\rm A}(z)$ and $D_{\rm B}(z)$ given by \citet{Madau95}.}
\label{fig:ext}
\end{figure}

The dust extinction effect is also included for the SEDs when generating the mock flux data. The flux density or SED after interstellar dust reddening can be expressed as \citep{Calzetti94,Galametz17}
\be \label{eq:SED_ext}
S_{\rm ext}(\lambda_{\rm res}) = S_{\rm int}(\lambda_{\rm res})\,10^{-0.4E(B-V)k(\lambda_{\rm res})},
\ee
Here $E(B-V)=A_V/R_V$ is the color excess, $k(\lambda)$ is the dust extinction curve which is shown in Figure \ref{fig:ext}. We consider five extinction laws here, which are derived from the studies of the Milky Way (MW) \citep{Allen76,Seaton79}, Large Magellanic Cloud (LMC) \citep{Fitzpatrick86}, Small Magellanic Cloud (SMC) \citep{Prevot84,Bouchet85}, and starburst galaxy (SBG) \citep{Calzetti00}. The $R_V$ for these laws are 3.1, 3.1, 3.1, 2.72, and 4.05, respectively. For each galaxy source, the value of $E(B-V)$ and corresponding extinction law is given by the COSMOS catalog, and we directly use them to generate our mock flux data.


In addition to the extinction from interstellar dust in galaxies, we also consider the extinction due to absorption of intergalactic medium (IGM) for high-$z$ galaxies. When propagating from high-$z$ galaxies to the observer, the emission at shorter wavelength than the Ly$\alpha$ line can be absorbed by neutral hydrogen clouds in the IGM. In order to include this effect, we make use of the attenuation laws computed by \citet{Madau95}. They give the average flux decrements $D_{\rm A}$ between Ly$\alpha$ and Ly$\beta$, and $D_{B}$ between Ly$\beta$ and the Lyman limit \citep{Oke82,Schneider91}, which are defined as
\be
D_{\rm i} \equiv \langle 1-S_{\rm abs}/S_{\rm ini} \rangle, \ \ {\rm i=A, B}.
\ee
Here $S_{\rm abs}$ is the flux density after IGM absorption, and $S_{\rm ini}=S_{\rm ext}$ is the initial flux density in the rest frame including interstellar dust reddening given by Equation~(\ref{eq:SED_ext}). In Figure \ref{fig:ext}, we show the $D_{\rm A}$ and $D_{\rm B}$ as a function of redshift. After obtaining $S_{\rm abs}$,  we get the $S_{\rm model}$ in Equation~(\ref{eq:f_obs}) for generating mock flux data.

\begin{figure}
\includegraphics[width=1.05\columnwidth]{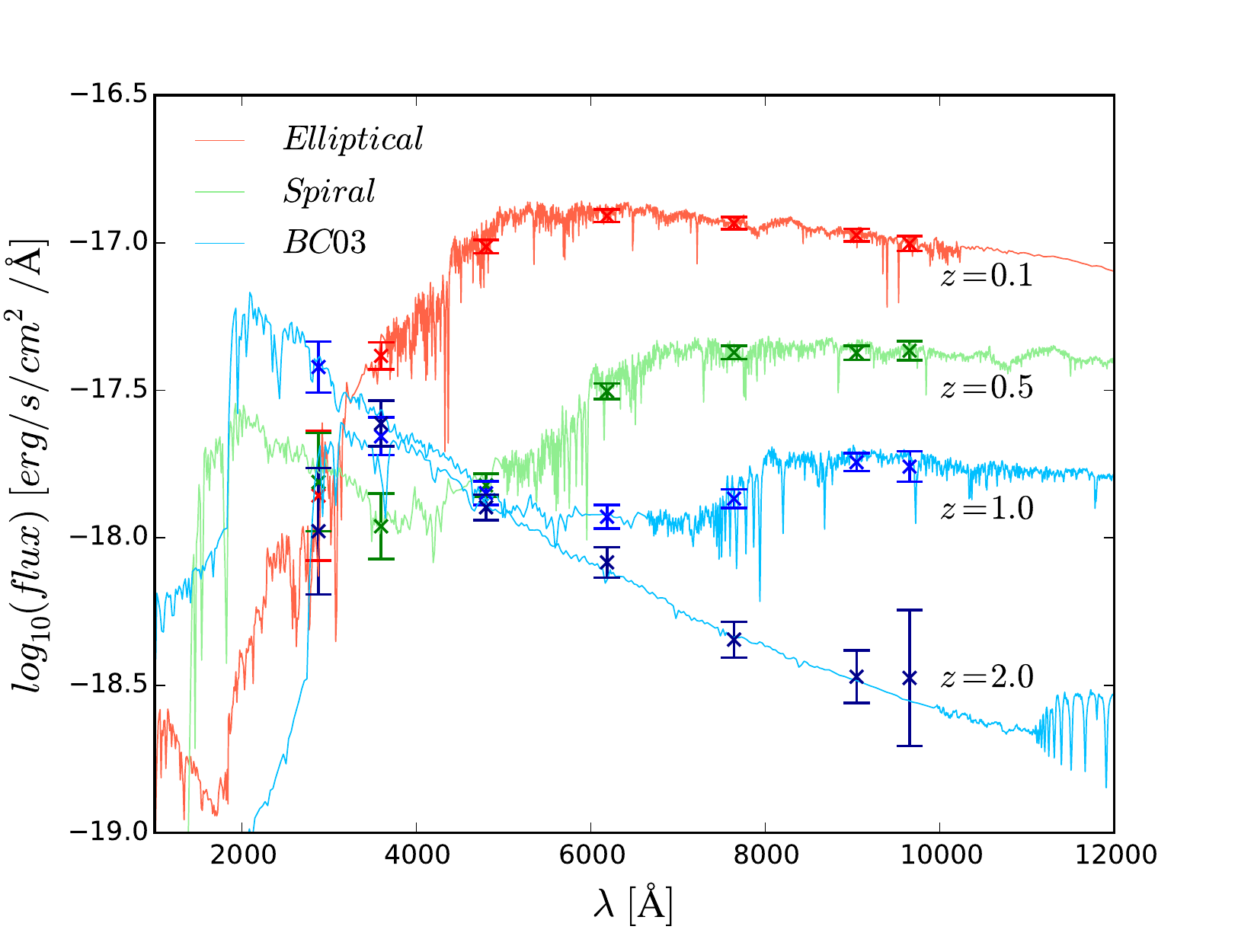}
\caption{ Examples of mock flux data of four randomly selected galaxies for the seven bands of the CSS-OS at $z=0.1$, $0.5$, $1.0$, and $2.0$. The curves denote the mock observed SED, and the crosses with error bars are the mock flux data.  A random error drawn from Gaussian distribution function with $\sigma=\sigma_F$ is included in the mock photometry.}
\label{fig:flux}
\end{figure}

We now estimate the flux error measured by the CSS-OS. For a space telescope, the signal to noise ratio (SNR) can be evaluated by \citet{Ubeda11}
\be \label{eq:SNR}
{\rm SNR} = \frac{C_{\rm s}t}{\sqrt{C_{\rm s}t+N_{\rm pix}(B_{\rm sky}+B_{\rm det})\,t+N_{\rm pix}N_{\rm read}R_{\rm n}^2}},
\ee
where $t$ is the exposure time, $N_{\rm pix}$ is the number of detector pixels covered by an object for the CSS-OS, which can be derived from the values given by the COSMOS catalog, $N_{\rm read}$ is the number of detector readouts, $B_{\rm det}$ is the detector dark current, and $R_{\rm n}$ is the read noise. In the CSS-OS, we set $t=300$ s, $N_{\rm read}=2$, $B_{\rm det}=0.02\ e^-{\rm /s}/{\rm pixel}$, and $R_{\rm n}=5\ e^-/{\rm pixel}$. $C_{\rm s}$ is the count rate from the source in $e^-/\rm s$, which can be calculated by
\be
C_{\rm s} = A_{\rm eff} \int S_{\rm obs}(\lambda) \tau(\lambda) \frac{\lambda}{hc} d\lambda,
\ee 
where $A_{\rm eff}$ is the effective aperture area of the telescope that varies for different bands, and $h$ and $c$ are the Planck constant and speed of light, respectively. The system throughput $\tau(\lambda)=m_{\rm eff}(\lambda)T_{\rm filter}(\lambda)$, where $m_{\rm eff}(\lambda)$ is the mirror efficiency, and $T_{\rm filter}(\lambda)$ is filter transmission. The CSS-OS mirror efficiency is found to be $\sim$0.5 for $NUV$ band, $\sim$0.7 for $u$ band, and $\sim$0.8 for other bands. $B_{\rm sky}$ in Eq. (\ref{eq:SNR}) is the sky background in $e^-/{\rm s}/{\rm pixel}$, which is given by
\be
B_{\rm sky} = A_{\rm eff} \int I_{\rm sky}(\lambda)\, l^2_{\rm p}\, \tau(\lambda) \frac{\lambda}{hc} d\lambda,
\ee
where $I_{\rm sky}$ is the surface brightness of the sky background in $\rm erg\, s^{-1}cm^{-2}\AA^{-1}arcsec^{-2}$, and $l_{\rm p}$ is the detector pixel scale. $B_{\rm sky}$ depends on many components, such as the zodiacal light, earthshine, phase of moon, etc. Here we estimate $B_{\rm sky}$ based on the throughput of the CSS-OS filters and the measurements of the zodiacal light and earthshine for ``average" sky background case given in \citet{Ubeda11}. We find that $B_{\rm sky}$ are 0.003, 0.018, 0.156, 0.200, 0.207, 0.123 and 0.036 $e^-/{\rm s}/{\rm pixel}$ for $NUV$, $u$, $g$, $r$, $i$, $z$ and $y$ band, respectively. 


The photometric error can be evaluated by the magnitude error given by the approximate relation $\sigma_{\rm ph}\simeq 2.5\,{\rm log_{10}}\left[ 1+1/{\rm SNR}\right]$ \citep{Bolzonella00,Pozzetti96,Pozzetti98}. We also add a systematic error $\sigma_{\rm sys}=0.02$ mag for all observational data, and the total magnitude error is then given by $\sigma_m=\sqrt{\sigma_{\rm ph}^2+\sigma_{\rm sys}^2}$. Then we can find the flux error $\sigma_F$ for each band from $\sigma_m$ via error propagation. 
In Figure \ref{fig:flux}, examples of mock flux data of four randomly selected galaxies are shown at $z=0.1$, $0.5$, $1.0$, and $2.0$. Finally, to each mock flux, a random error drawn from Gaussian probability distribution function (with $\sigma = \sigma_{F}$) is added in the mock photometry.

\section{Photo-z code}

In this section, we test three widely used photo-$z$ template-fitting codes, i.e. LePhare\footnote{\tt http://www.cfht.hawaii.edu/~arnouts/LEPHARE/lephare.html}\citep{Arnouts99,Ilbert06}, EAZY\footnote{\tt http://www.astro.yale.edu/eazy/}\citep{Brammer08}, and Hyperz\footnote{\tt http://webast.ast.obs-mip.fr/hyperz/}\citep{Bolzonella00}. Using the galaxy catalog and the mock flux data described in the last section, we fit the photometric redshift  with these three codes, respectively. In order to compare the fitting results, we use the same SED templates provided by LePhare for all these three codes. In these codes, the photometric redshift is derived by the least-square method, which minimizes the following chi-square:
\be \label{eq:chi_N}
\chi^2 = \sum_{i}^{N}\left( \frac{F_i^{\rm obs}-F_i^{\rm th}}{\sigma_i^{\rm obs}} \right)^2,
\ee
where $N$ is the number of bands, $F_i^{\rm obs}$ and $\sigma_i^{\rm obs}$ are the observed flux and error for band $i$, respectively, which are derived from the mock flux data in the last section. $F_i^{\rm th}$ is the predicted flux by photo-$z$ fitting code. 

\begin{figure*}
\includegraphics[width=0.69\columnwidth]{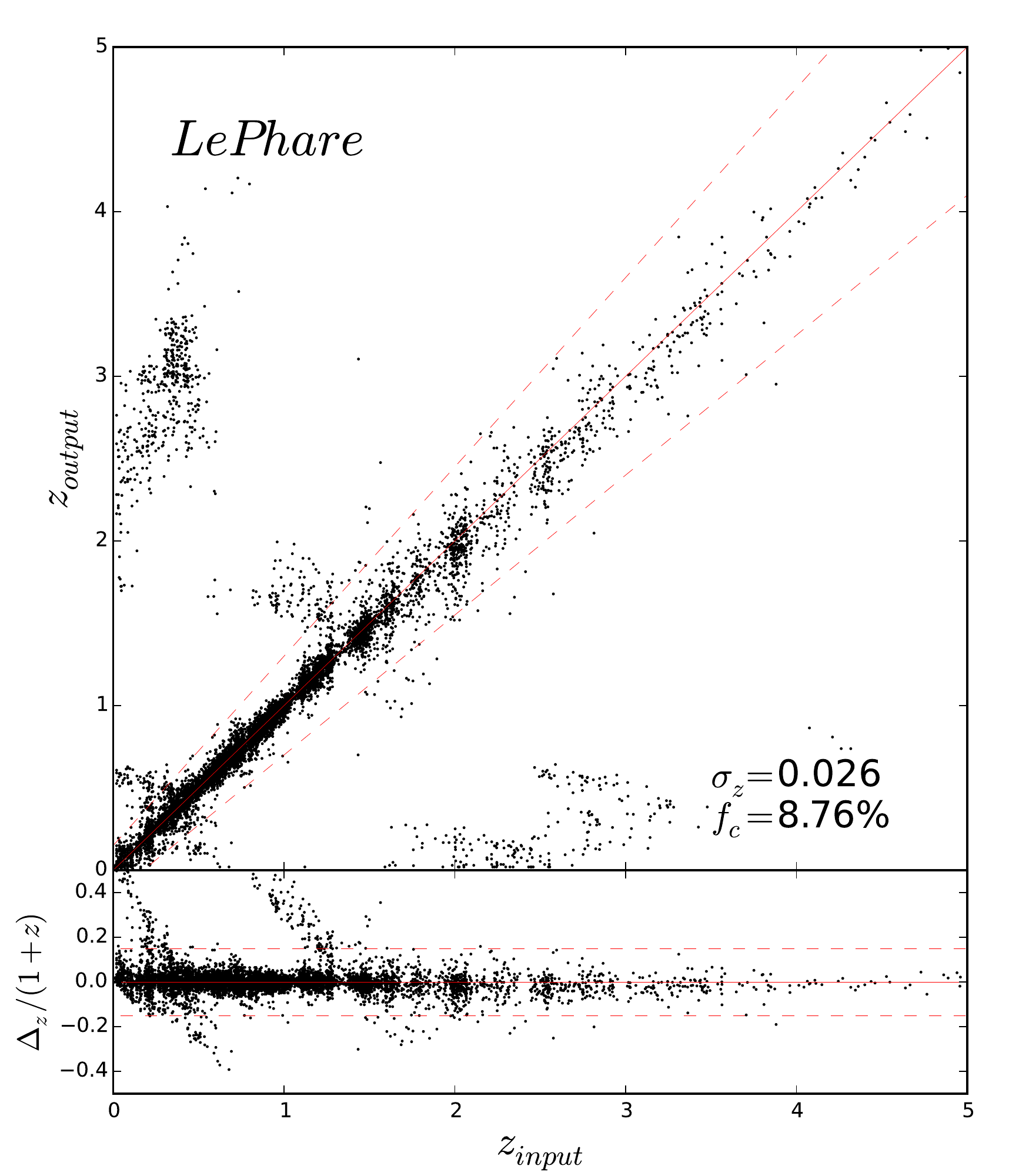}
\includegraphics[width=0.69\columnwidth]{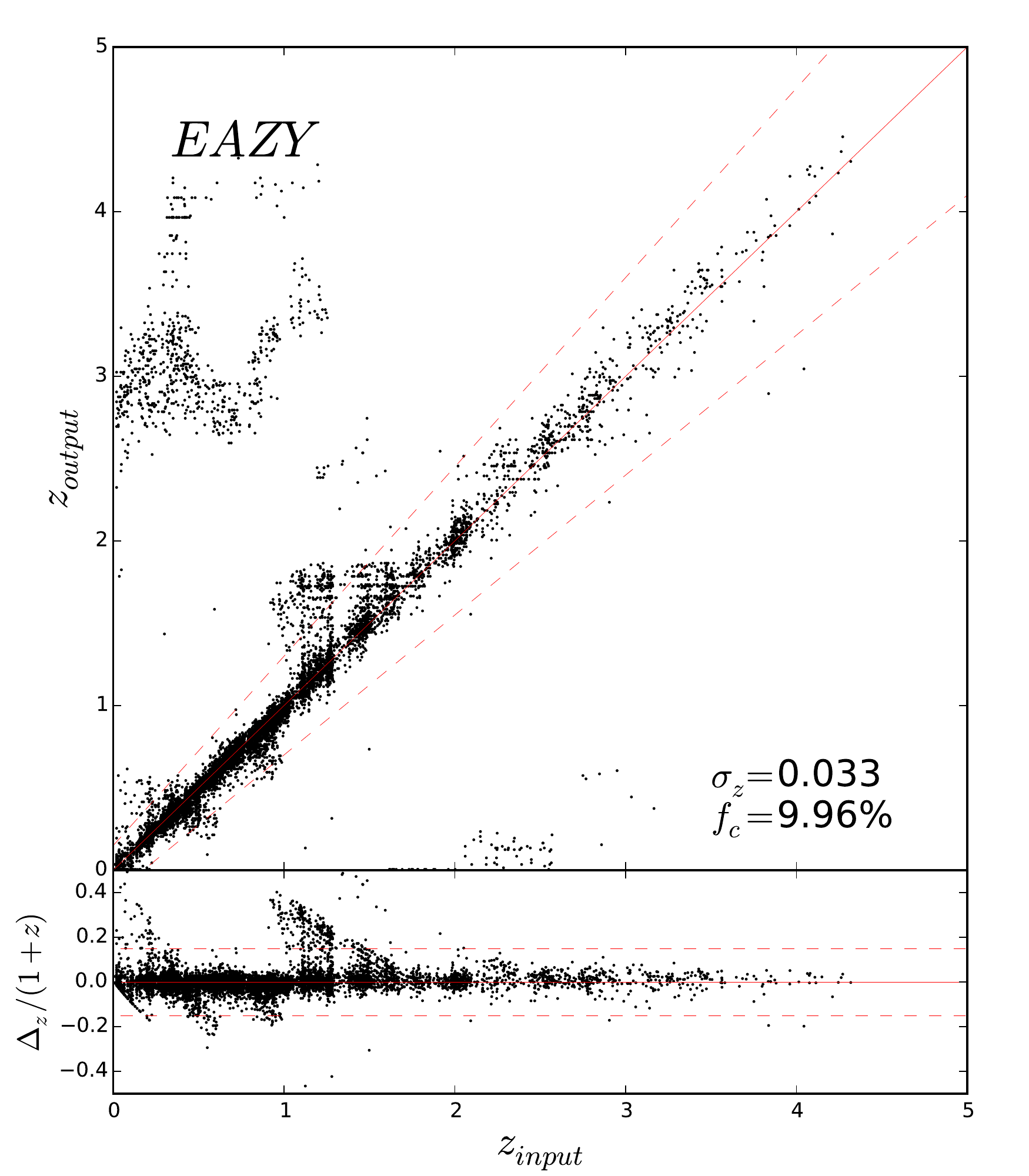}
\includegraphics[width=0.69\columnwidth]{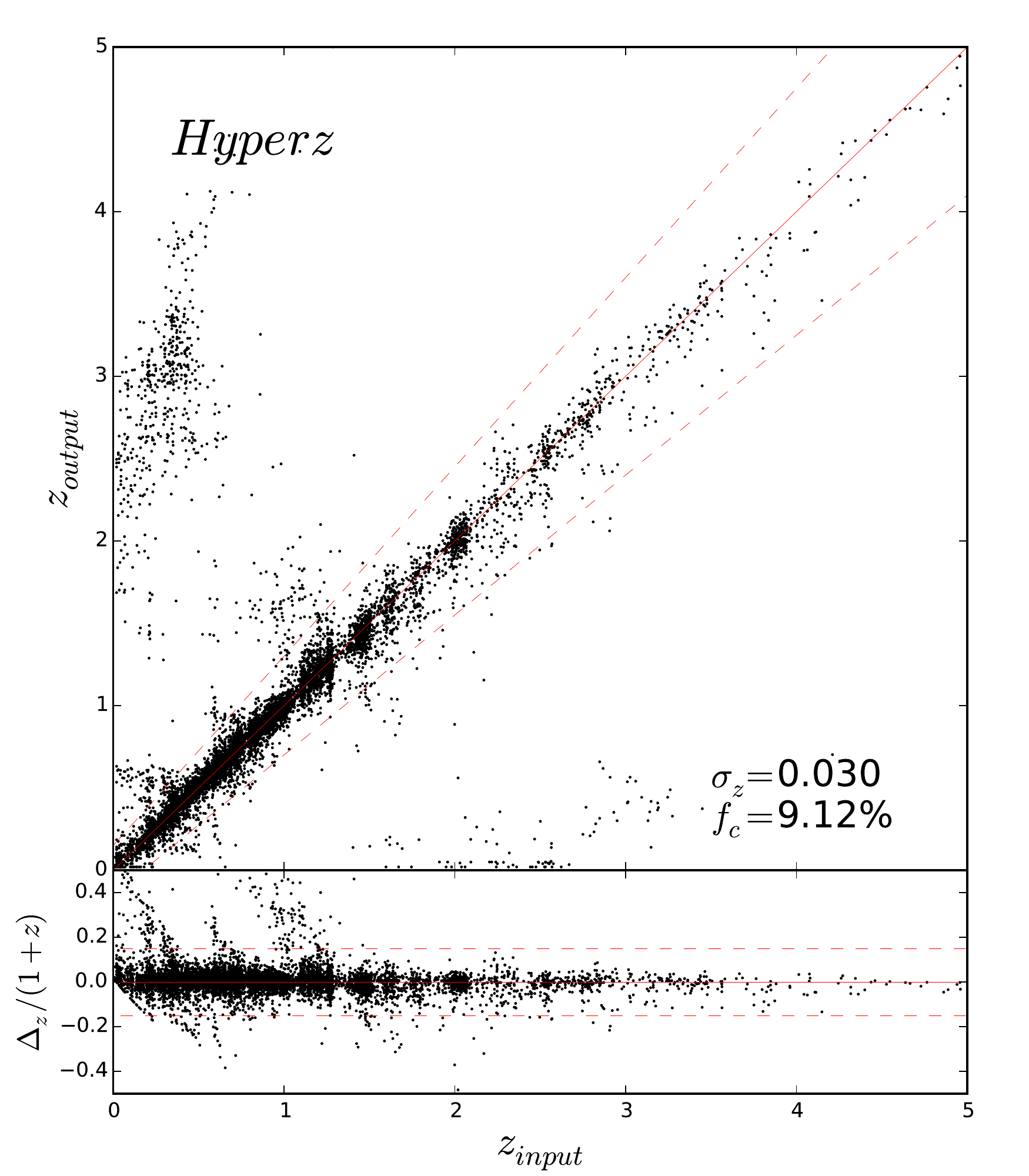}
\caption{The photometric redshift fitted by the LePhare, EAZY and Hyperz codes excluding the bands with flux information in upper-limits. We find that LePhare gives $\sigma_z=0.026$ and $f_{\rm c}=8.76\%$, which is the most accurate photo-$z$ result. For comparison, EAZY and Hyperz give $\sigma_z=0.033$ and $f_{\rm c}=9.96\%$, and $\sigma_z=0.030$ and $f_{\rm c}=9.12\%$, respectively. The red dashed lines indicate the bounds of the catastrophic redshift identification, which is defined as $|\Delta z|/(1+z_{\rm input})>0.15$.}
\label{fig:zp_code}
\end{figure*}

In Figure {\ref{fig:zp_code}}, we show $z_{\rm input}$ vs. $z_{\rm output}$ for the three photo-$z$ codes. 
Here $\Delta z=z_{\rm output}-z_{\rm input}$, where $z_{\rm input}$ and $z_{\rm output}$ are input and output redshifts, respectively, $\sigma_z$ is the total deviation of the photo-$z$ fitting. Here we take the normalized median absolute deviation (NMAD) \citep{Brammer08} in the calculation, and it is given by
\be
\sigma_{\rm NMAD} = 1.48 \times {\rm median}\left( \left| \frac{\Delta z-{\rm median}(\Delta z)}{1+z_{\rm input}}\right|\right).
\ee 
The advantage of this deviation is that it can naturally suppress the weighting of catastrophic redshift identifications \citep{Ilbert06}, which is defined as $|\Delta z|/(1+z_{\rm input})>0.15$ here, and give a proper estimation of the total photo-$z$ accuracy. We find that $\sigma_z=\sigma_{\rm NMAD}=0.026$ and the catastrophic redshift fraction $f_{\rm c}=8.76\%$ for LePhare, which is the most accurate photo-$z$ result. The results of Hyperz are $\sigma_z=0.030$ and $f_{\rm c}=9.12\%$, and they are greater than the results of LePhare. EAZY gives $\sigma_z=0.033$ and $f_{\rm c}=9.96\%$, which are higher than the results of both LePhare and Hyperz. Hereafter, we would use the LePhare code to fit photo-$z$ and perform the filter calibration.

In the photo-$z$ fitting process discussed above, we discard data points in bands where they are too low to be detected above 3$\sigma$ sensitivity limits. Obviously, this can lead to information loss for the bands with low detection efficiency, especially for the $NUV$, $u$, $z$ and $y$ bands as shown in the right panel of Figure \ref{fig:filters}. However, the data in these four bands are quite valuable for distinguishing the Lyman and Balmer break features of galaxy SEDs, and can suppress the fraction of catastrophic redshift.

In order to exploit information in upper-limits, we make use of the following total $\chi^2$ in the estimation, which can be expressed as
\be \label{eq:chi2_tot}
\chi^2_{\rm tot} = \chi^2_N + \sum_j^M w_j,
\ee
where $\chi^2_N$ is for the data with SNR$\ge$3 as shown in Eq.~(\ref{eq:chi_N}), $M$ is the number of  the bands with upper-limits, and $w_j=-2\,{\rm log}P_j$. Here $P_j$ is given by
\be \label{eq:Pj}
P_j = \frac{1}{\sqrt{2\pi} \sigma_j}\int_{F_{\rm l}}^{F_{\rm u}} {\rm exp}\left[ -\frac{(f-F_j^{\rm th})^2}{2\sigma_j^2}\right] df,
\ee
where $f$ is the flux variable, $F_j^{\rm th}$ is the flux calculated by the photo-$z$ code for band $j$, $\sigma_j$ is the flux error of band $j$, $F_{\rm l}$ and $F_{\rm u}$ are the flux lower and upper limits, and we take $F_{\rm l}=0$ and $F_{\rm u}=3\,\sigma_j$. Such a $\chi^2$ enables a maximum-likelihood estimation of photo-z \citep{Isobe86,Lyu16}. Note that we set $F_{\rm l}=0$ instead of $-\infty$, since the UV and optical backgrounds are quite low. As we discuss below, this method can naturally include the information in upper-limits, and efficiently suppress the catastrophic redshift fraction.

\begin{figure*}
\centerline{
\resizebox{!}{!}{\includegraphics[scale=0.42]{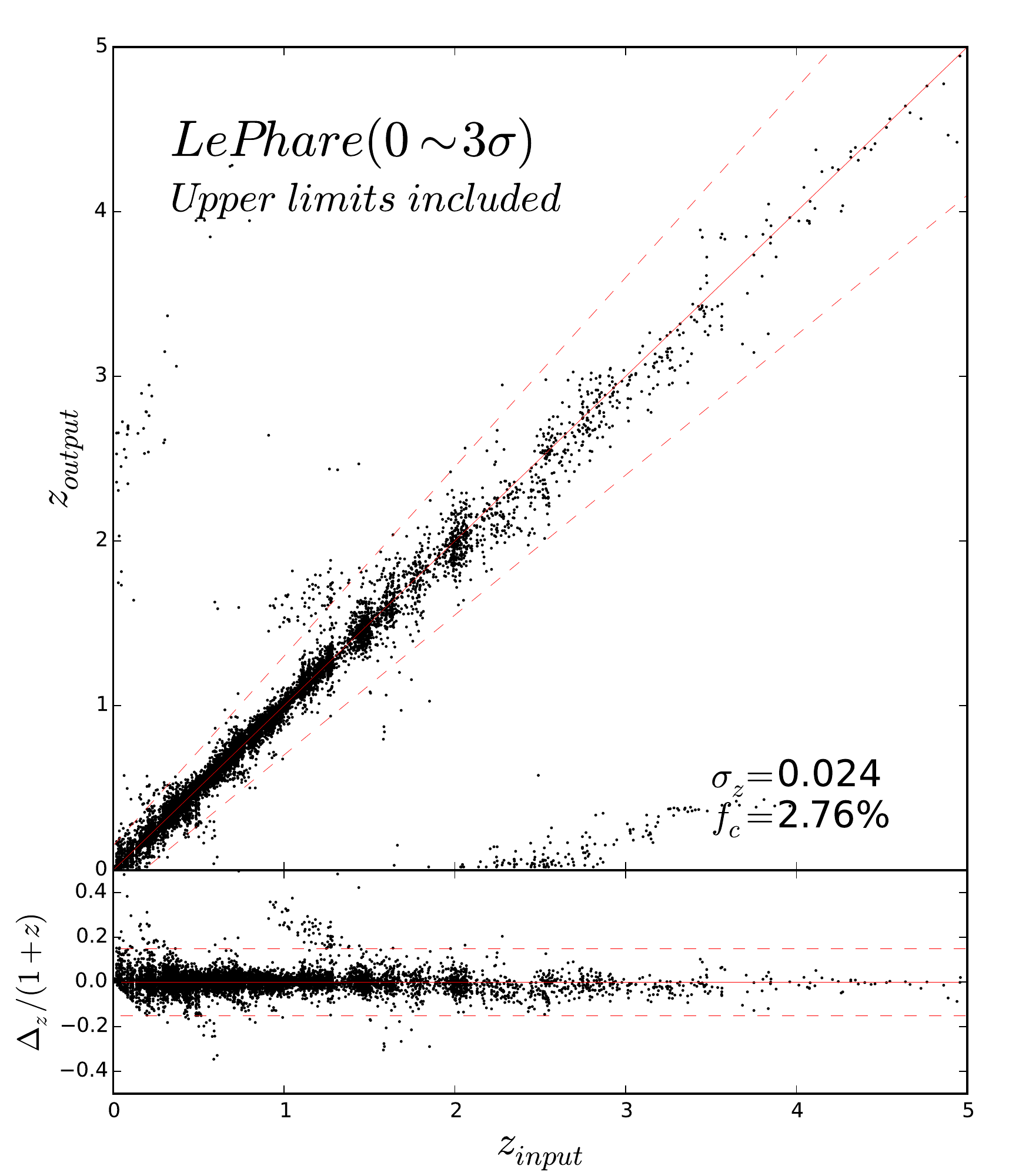}}
\resizebox{!}{!}{\includegraphics[scale=0.42]{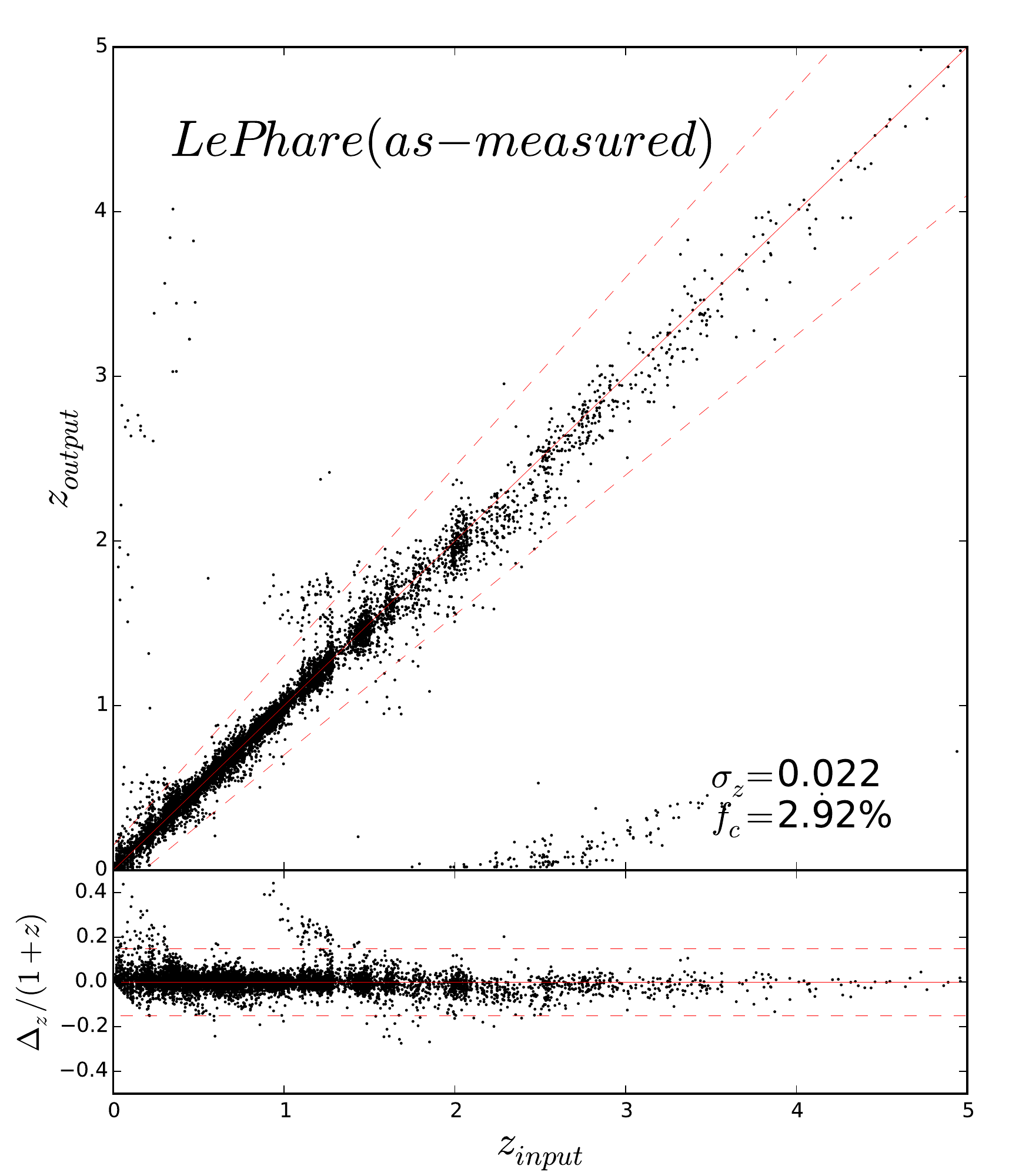}}
}
\caption{\label{fig:LePhare_mod} $Left:$ The photo-$z$ fitting results including the information in upper-limits using Eq.~(\ref{eq:chi2_tot}). Although $\sigma_z$ is similar, the catastrophic redshift fraction $f_{\rm c}=2.76\%$, which is significantly improved compared to $f_{\rm c}=8.76\%$ without considering the information of upper limits as shown in the left panel of Figure \ref{fig:zp_code}. $Right:$ The results of using as-measured flux and error even for the data with sensitivity below $3\sigma$. As can be seen, these two methods can provide similar result.}
\end{figure*}

We modify the LePhare code by replacing Eq.~(\ref{eq:chi_N}) with Eq.~(\ref{eq:chi2_tot}). The results are presented in the left panel of  Figure {\ref{fig:LePhare_mod}}. We find that $\sigma_z=0.024$ is close to the result without including the upper limits, since our $\sigma_z$ is insensitive to the catastrophic redshift. On the other hand, the current catastrophic redshift fraction $f_{\rm c}=2.76\%$, which is significantly improved compared to $f_{\rm c}=8.76\%$ given by the case without including the upper limits. By comparing the left panels of Figure~\ref{fig:LePhare_mod} and Figure~\ref{fig:zp_code}, we can see that the number of the poorly fitted dots around $z_{\rm input}=0.3$, which are caused by the misidentification of Lyman and Balmer breaks in SEDs, is remarkably suppressed. Hence, our method can properly take into account of the information in upper-limits, and significantly reduce the catastrophic redshift fraction. We will adopt this method to include the bands with information in upper-limits in the following discussion of Section 4.

Alternatively, another method can be adopted, which directly uses the ``as-measured" flux and error with Eq.~(\ref{eq:chi_N}), even for the data with sensitivity below $3\sigma$. As shown in the right panel of Figure~\ref{fig:LePhare_mod}, we find that $\sigma_z=0.022$ and $f_c=2.92\%$, which is similar to the result in the left panel. Hence, both methods can be used in the CSS-OS photo-$z$ fitting process. In addition, we also compare other three methods in Appendix.

\section{Photo-z dependency}

After obtaining mock flux data, we use the modified LePhare code to perform photo-$z$ calibration and test the filter definition. We first explore the effect of each filter passband on photo-$z$ accuracy by removing it. We then investigate if the current CSS-OS filter definition can provide accurate photo-$z$ results that can achieve the science requirement in certain filter parameter ranges. We also compare the CSS-OS filter set to other types of filters.

\subsection{Dependency of photo-z accuracy on each band}

\begin{figure*}
\includegraphics[width=\columnwidth]{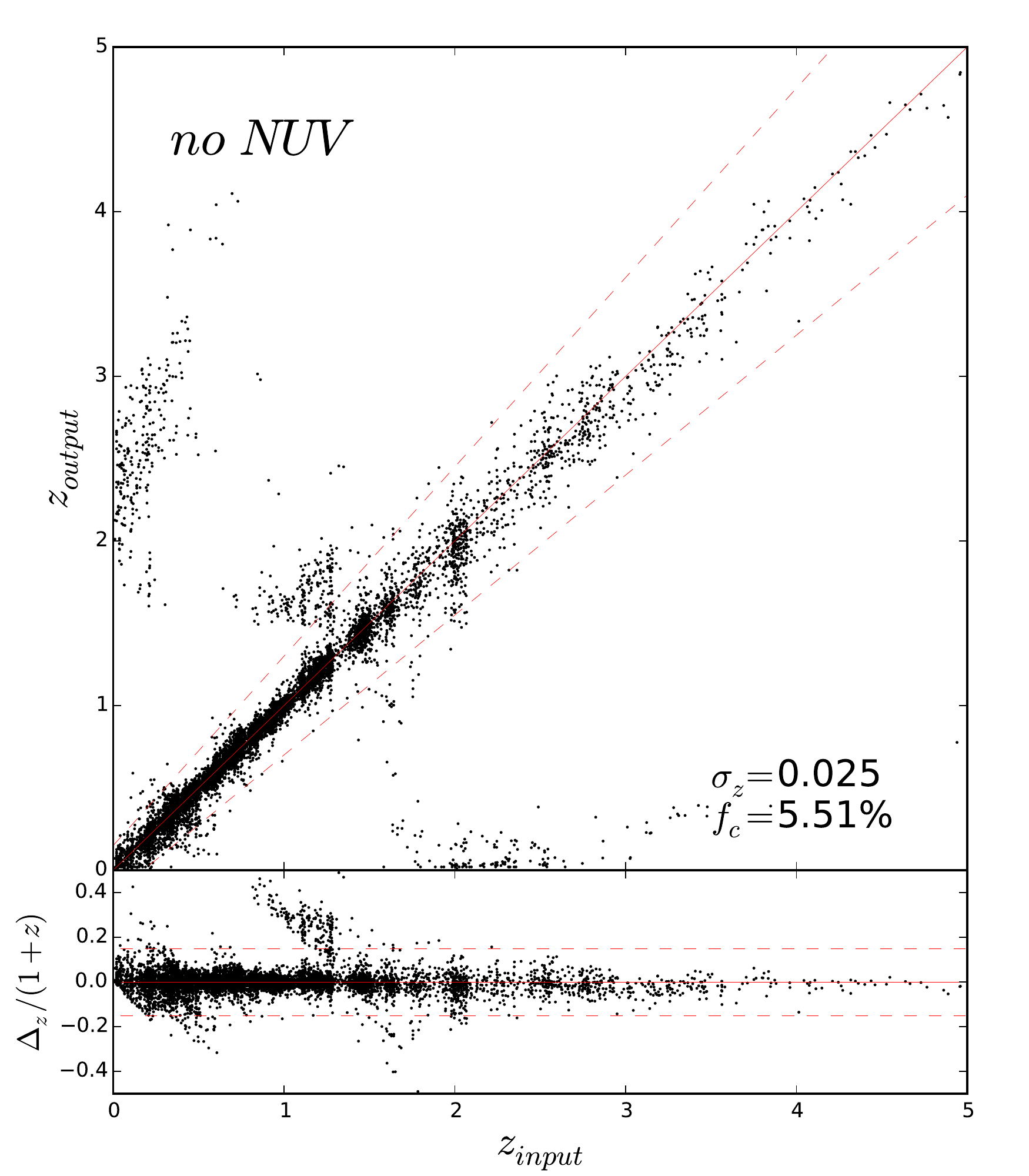}
\includegraphics[width=\columnwidth]{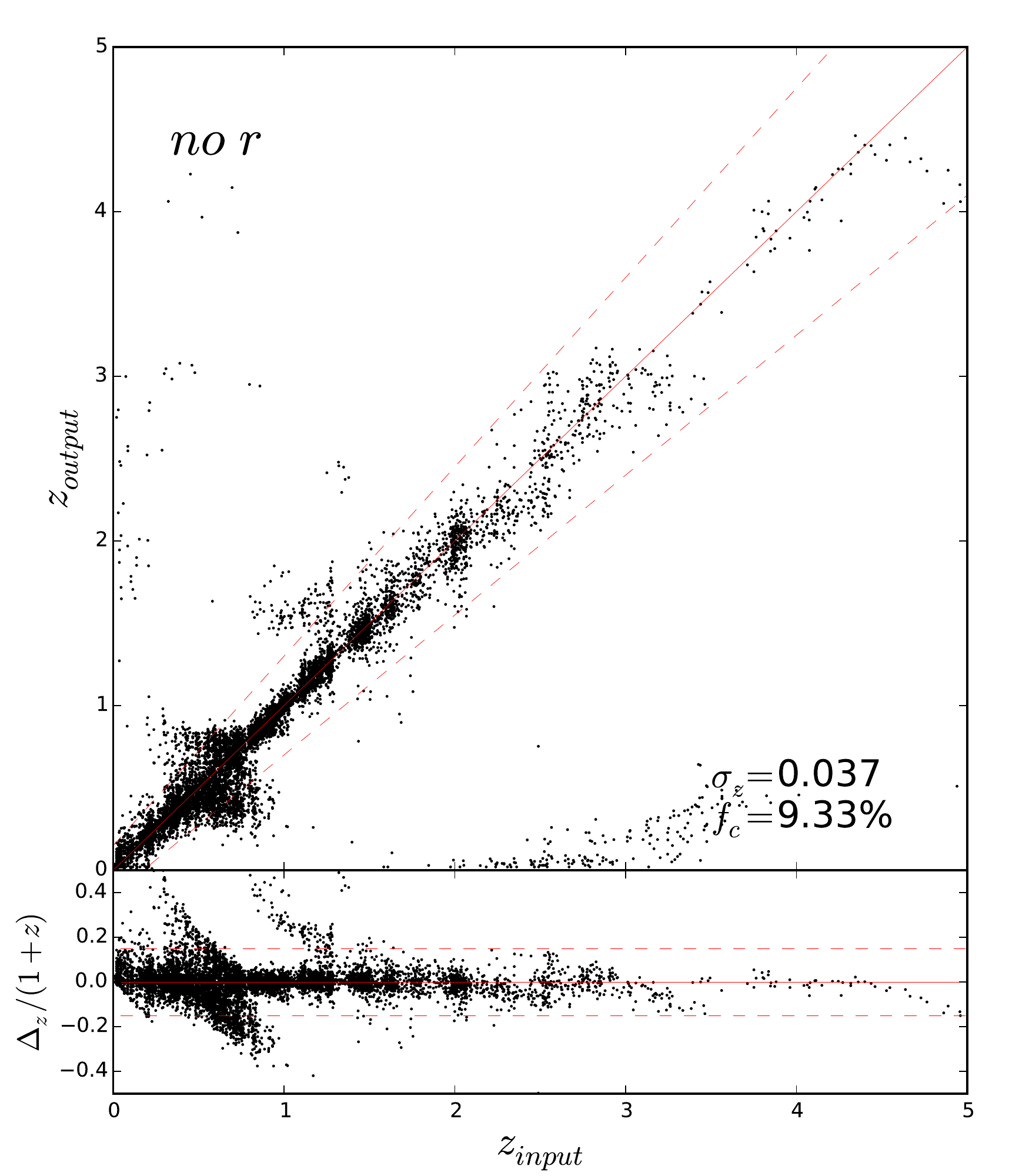}
\includegraphics[width=\columnwidth]{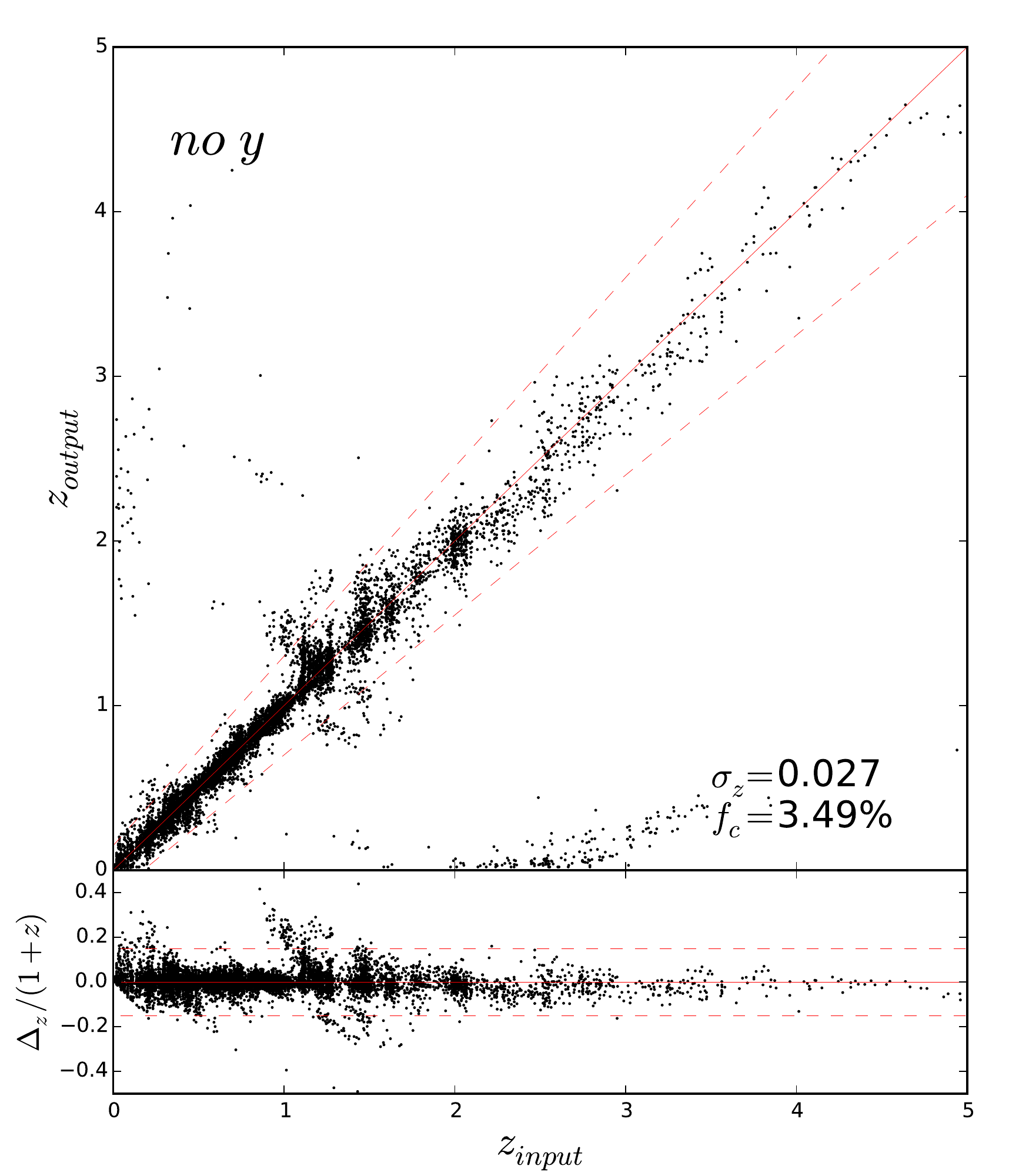}
\includegraphics[width=\columnwidth]{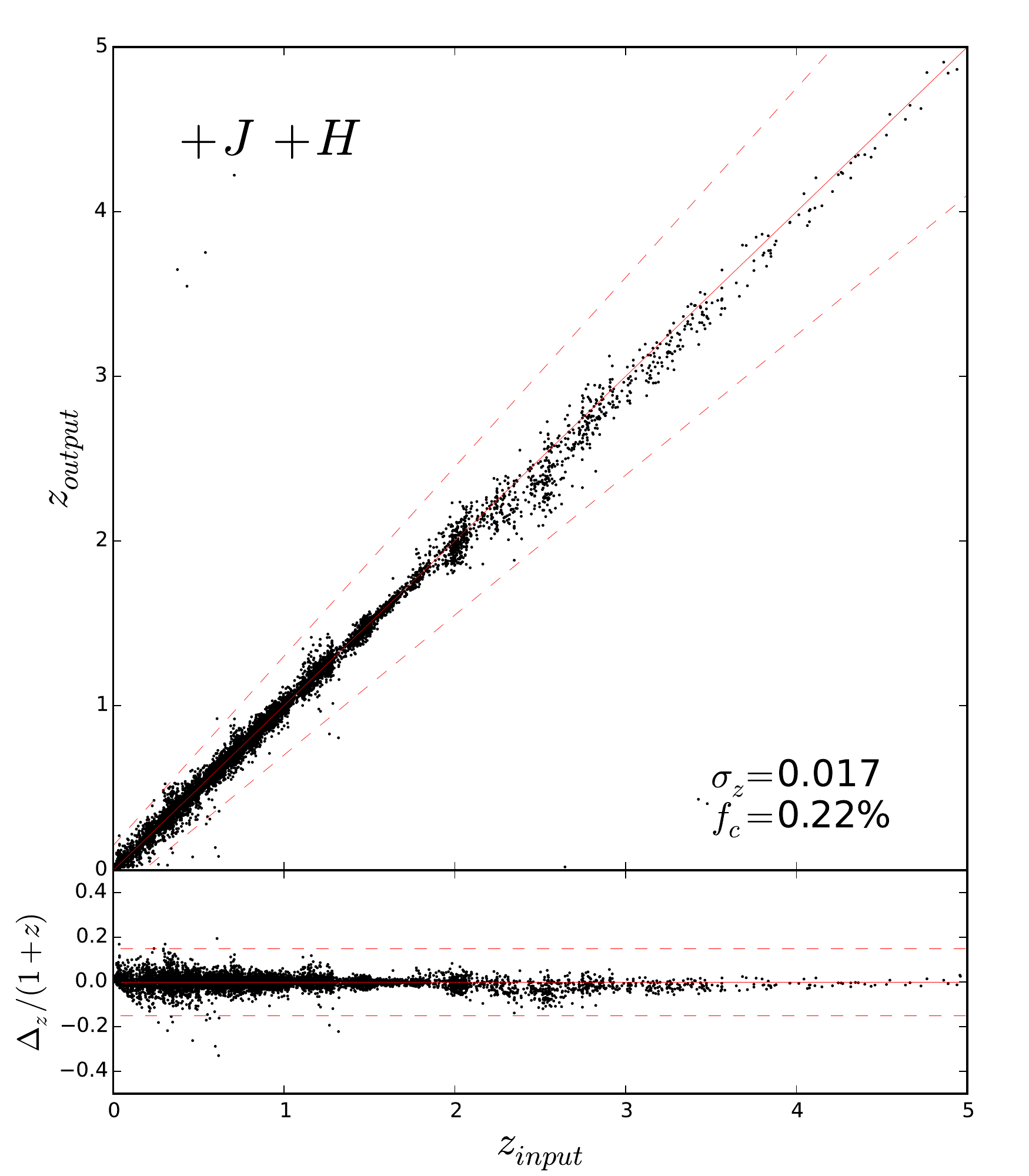}

\caption{The photo-$z$ fitting results for the seven bands without $NUV$,  $r$ and $y$, and with $J+H$ bands, respectively. We find that the $NUV$, $g$, $r$, and $i$ bands have large effects on the photo-$z$ results, and the missing of these bands will significantly suppress the photo-$z$ accuracy. On the other hand, the $u$, $z$ and $y$ bands don't affect the results as much as the other four bands in our fittings. The $J$ and $H$ bands at longer wavelengths can improve the fitting results, especially for the objects at $z>$3. Note that the results of ``no $NUV$'', ``no $r$", and ``no $y$" are assuming the real detector efficiency, and it is 100\% efficiency for the result of the ``$+J+H$" case.}
\label{fig:filter_set}
\end{figure*}

In Figure~\ref{fig:filter_set}, we show the photo-$z$ fitting results for the seven filter bands in the CSS-OS survey that removing the $NUV$, $r$ and $y$ bands, and including additional $J$ and $H$ bands at NIR, respectively. The $\sigma_z$ and $f_{\rm c}$ results of removing each single band are shown in Table~\ref{tab:filter_zp}. In order to generally study the synergy with other surveys, we also include 100\% detector efficiency case. Since the transmission efficiencies are higher for all seven bands, we find that the results of 100\% detector efficiency are generally better than that of the real efficiency case. 

\begin{table}
\centering
\caption{The redshift fitting variance $\sigma_z$ and catastrophic redshift fraction $f_{\rm c}$ for different filter sets.}
\begin{tabular}{c c c c c}
\hline\hline
&\multicolumn{2}{c}{$\rm real\ det.\ eff.$}&\multicolumn{2}{c}{$\rm 100\%\ det.\ eff.$}\\
Filter set  & $\sigma_z$ & $f_{\rm c}(\%)$ & $\sigma_z$ &  $f_{\rm c}(\%)$ \\
\hline
 $All$ & 0.024 & 2.76 & 0.021  & 0.82 \\
 \hline
 $-NUV$ & 0.025  & 5.51 & 0.022  & 3.07 \\
 $-u$ & 0.026  & 3.99 & 0.022  & 1.44 \\
 $-g$ & 0.033  & 7.53 & 0.028  & 4.26 \\
 $-r$ & 0.037  & 9.33 & 0.030  & 4.60 \\
 $-i$ & 0.036  & 5.69 & 0.030  & 2.96 \\
 $-z$ & 0.028 & 3.44 & 0.023  & 1.65 \\
 $-y$ & 0.027 & 3.49 & 0.023  & 1.32 \\
\hline
 $+J$ & - & - & 0.017 & 0.43 \\
 $+J+H$ & - & - & 0.017 & 0.22 \\
\hline\end{tabular}
\label{tab:filter_zp}
\end{table}

When removing the $NUV$ band, as expected, we find that $\sigma_z=0.025$ and $f_{\rm c}=5.51\%$, which are higher than $\sigma_z=0.024$ and $f_{\rm c}=2.76\%$ in the case that it is included. The catastrophic redshift fraction $f_{\rm c}$ increases significantly, since we can see that there are many poorly fitted objects around $z=0.1$ in the top-left panel of Figure~\ref{fig:filter_set} compared to Figure~\ref{fig:LePhare_mod}. This is because that the continuum break at 2640 \AA\ can be misidentified as the Layman break without the $NUV$ band shown \cite[e.g.][]{Kriek11}. The effect of $u$ band is similar to that of the $NUV$ band, but has smaller impact on the fitting results. This indicates that the $NUV$ and $u$ band is important to pin down the number of catastrophic redshifts.

As shown in Table~\ref{tab:filter_zp}, the $g$, $r$ and $i$ bands can significantly affect the photo-$z$ fitting results. We have $\sigma_z=0.037$ and $f_{\rm c}\sim9\%$ if removing the $r$ band, that can decrease the photo-$z$ fitting accuracy dramatically. By checking the $z_{\rm input}$ vs. $z_{\rm output}$ results (e.g. see the top-right panel of Figure~\ref{fig:filter_set} for removing the $r$ band), we find that the $g$, $r$ and $i$ bands mainly affect the fittings at $z=0\sim0.5$, $0.3\sim0.8$, and $0.7\sim1.3$, respectively. The redshift dots in these ranges do not follow tightly the line $z_{\rm output}=z_{\rm input}$, but spread around it. It can be seen that these redshift ranges are around the peak of the galaxy distribution at $z\sim0.6$ shown in Figure~\ref{fig:zm_dis}, especially for the $r$ band, and thus can affect most galaxies observed by the CSS-OS. This is why these three bands are the most important for the CSS-OS. Without the $g$, $r$, and $i$ bands, we cannot identify both the Balmer break and 4000 \AA\ break in spectral features, and will give bad photo-$z$ estimates in the three redshift ranges mentioned above.

On the other hand, the $z$ and $y$ bands has smaller effects on the catastrophic redshift than the other five bands. Removing the $z$ or $y$ band can suppress the fitting accuracy at $z\sim1.5$ and $z\gtrsim3$ (see the bottom-left panel of Figure~\ref{fig:filter_set} for the effect of removing the $y$ band). However, they don't have large effects on the fitting results, since the number of objects in this redshift range is relatively small for the CSS-OS (see Figure~\ref{fig:zm_dis}).

In addition to the seven CSS-OS bands, to explore the synergy with other surveys, we also study the photo-$z$ calibration with other NIR bands, such as the $J$ and $H$ bands similar to  the Euclid telescope \citep{Laureijs11}. We set $\lambda_{-01}=1050$ nm, $\lambda_{+01}=1400$ nm, $\lambda_{-90}=1100$ nm, and $\lambda_{+90}=1350$ nm for the $J$ band, and $\lambda_{-01}=1350$ nm, $\lambda_{+01}=2000$ nm, $\lambda_{-90}=1400$ nm, and $\lambda_{+90}=1950$ nm for the $H$ band. The top transmission of these two bands is assumed to be 92\%, which is the same as that of the $i$, $z$, and $y$ bands. Then we perform photo-$z$ fitting using these two bands for the 100\% detector efficiency case. As we can see in Table~\ref{tab:filter_zp} and the bottom-right panel of Figure~\ref{fig:filter_set}, the photo-$z$ accuracy is improved by including these two bands. After comparing the plots of $z_{\rm input}$ vs. $z_{\rm output}$, we find that these two bands mainly can enhance the fitting accuracy at $z\sim2$ and $z>3$, since they are located at longer wavelengths \citep{Liu17}. Because there are not many objects in this redshift range for our survey, the overall improvement is limited. These two bands should be much more important in the surveys with higher magnitude limits, which can observe a lot of fainter and high-$z$ objects.

\subsection{Dependency of photo-z accuracy on filter parameters}

\begin{figure}
\centering
\vspace{-0.8cm}
\includegraphics[width=\columnwidth]{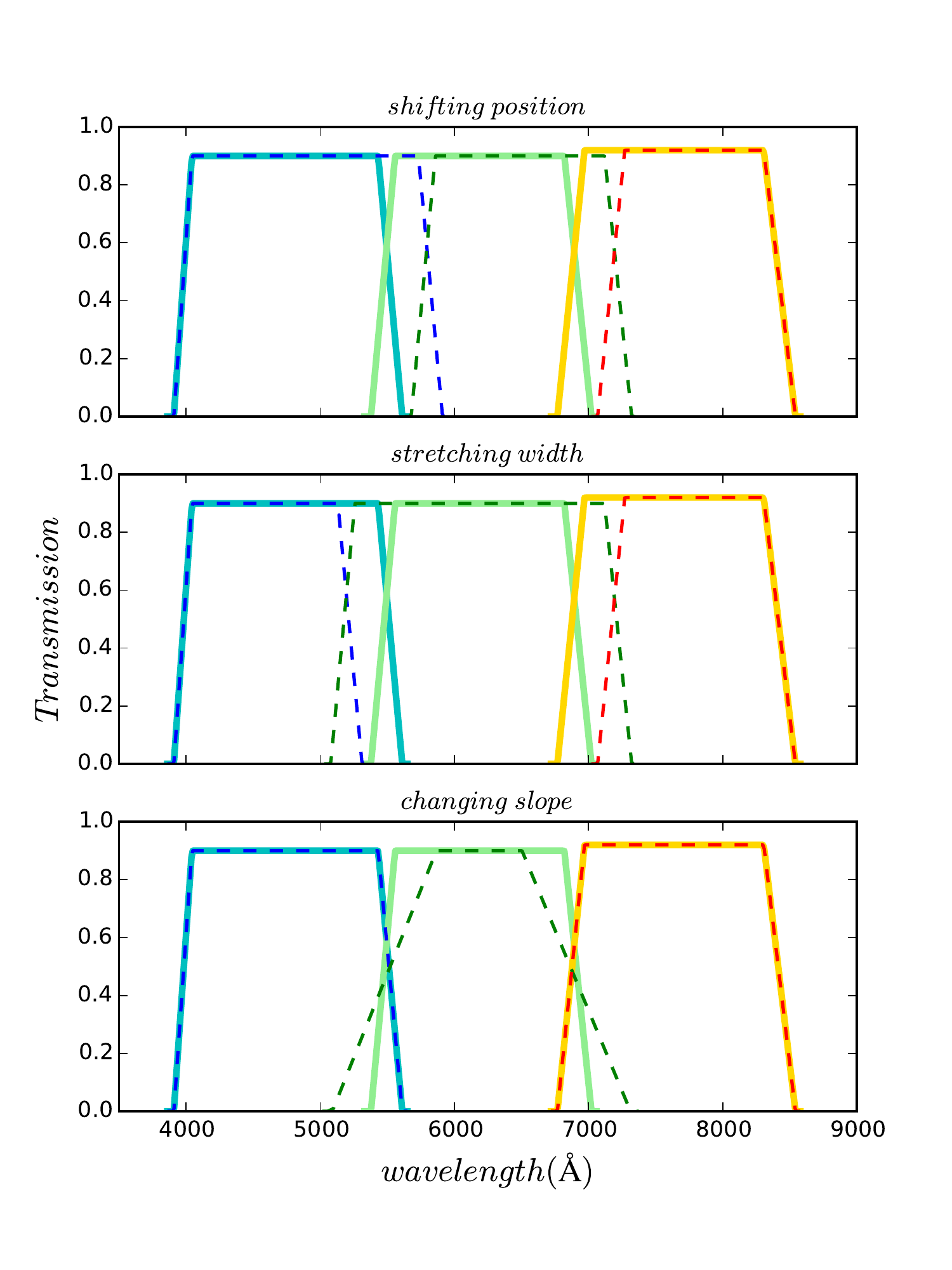}
\vspace{-1cm}
\caption{The diagrams of showing the strategy of changing the three filter parameters. The solid and dashed curves are the transmissions of the original and changed bands, respectively. When shifting or stretching the middle band (in green), we change the same wavelength for the adjacent bands (in blue and orange, other bands are fixed), in order to conserve the overlap regions between bands. For slope changing case (bottom panel, green curves), we fix the band FWHM, and change the top and bottom of the band simultaneously.}
\label{fig:cha_stra} 
\end{figure}

\begin{figure}
\centering
\includegraphics[width=0.87\columnwidth]{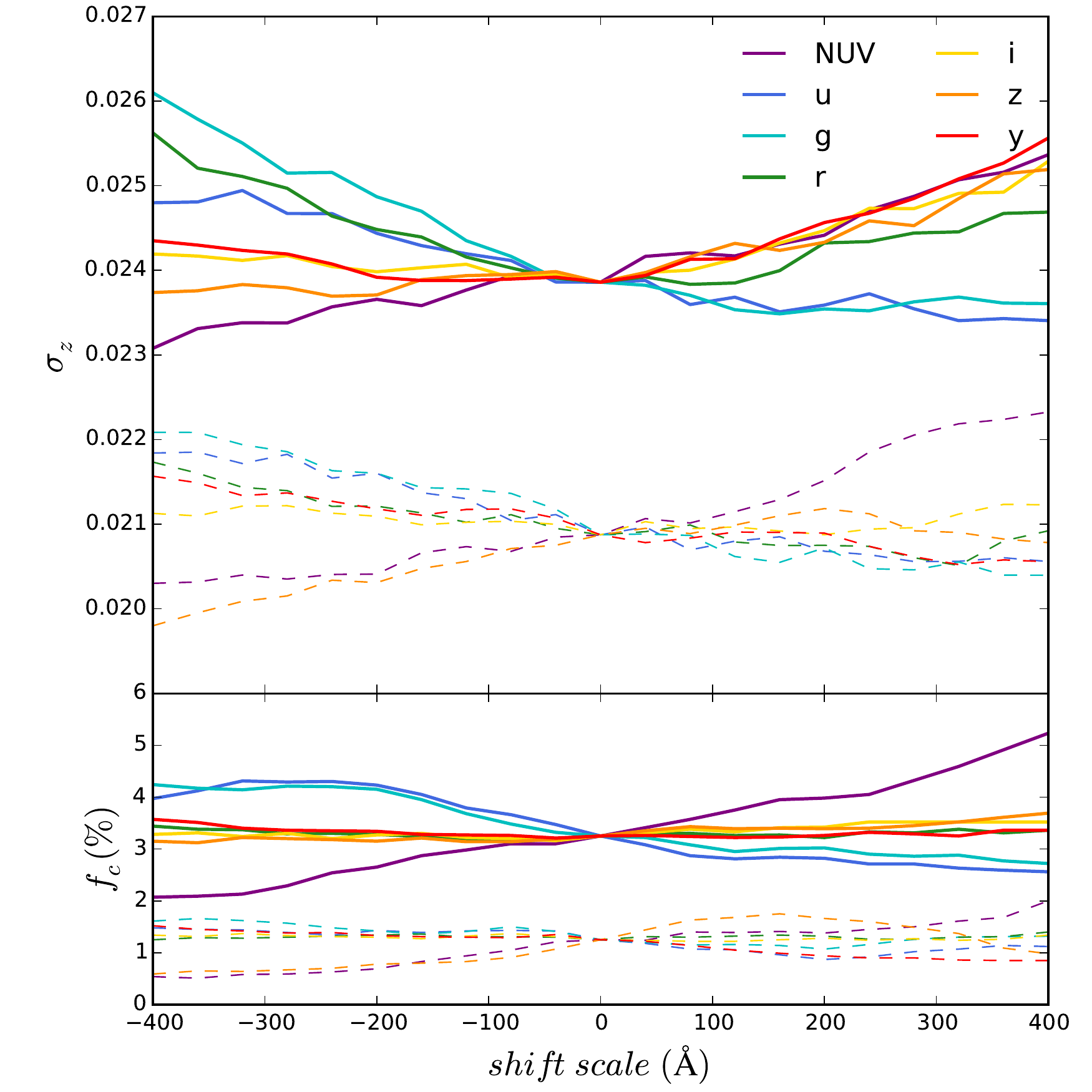}
\includegraphics[width=0.87\columnwidth]{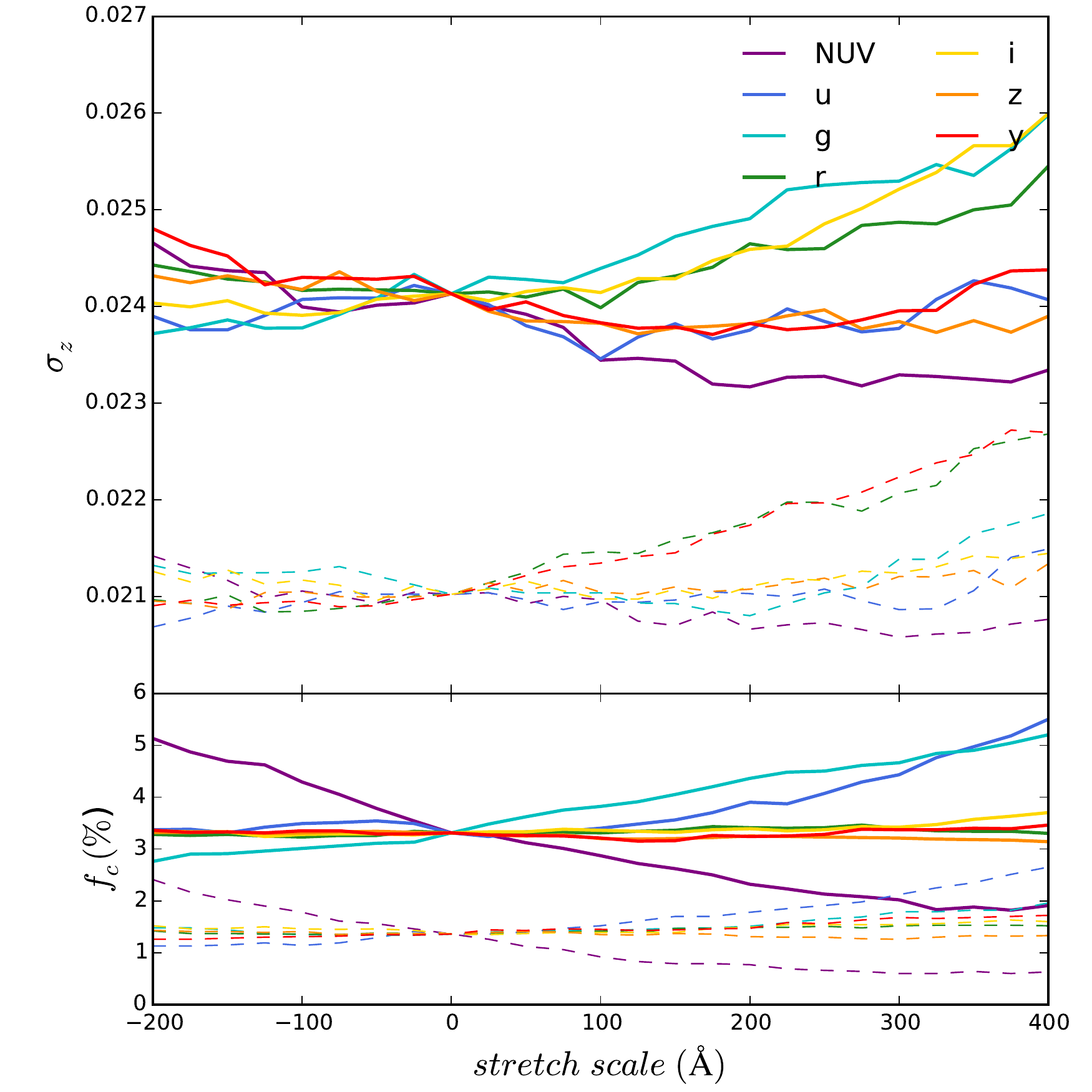}
\includegraphics[width=0.87\columnwidth]{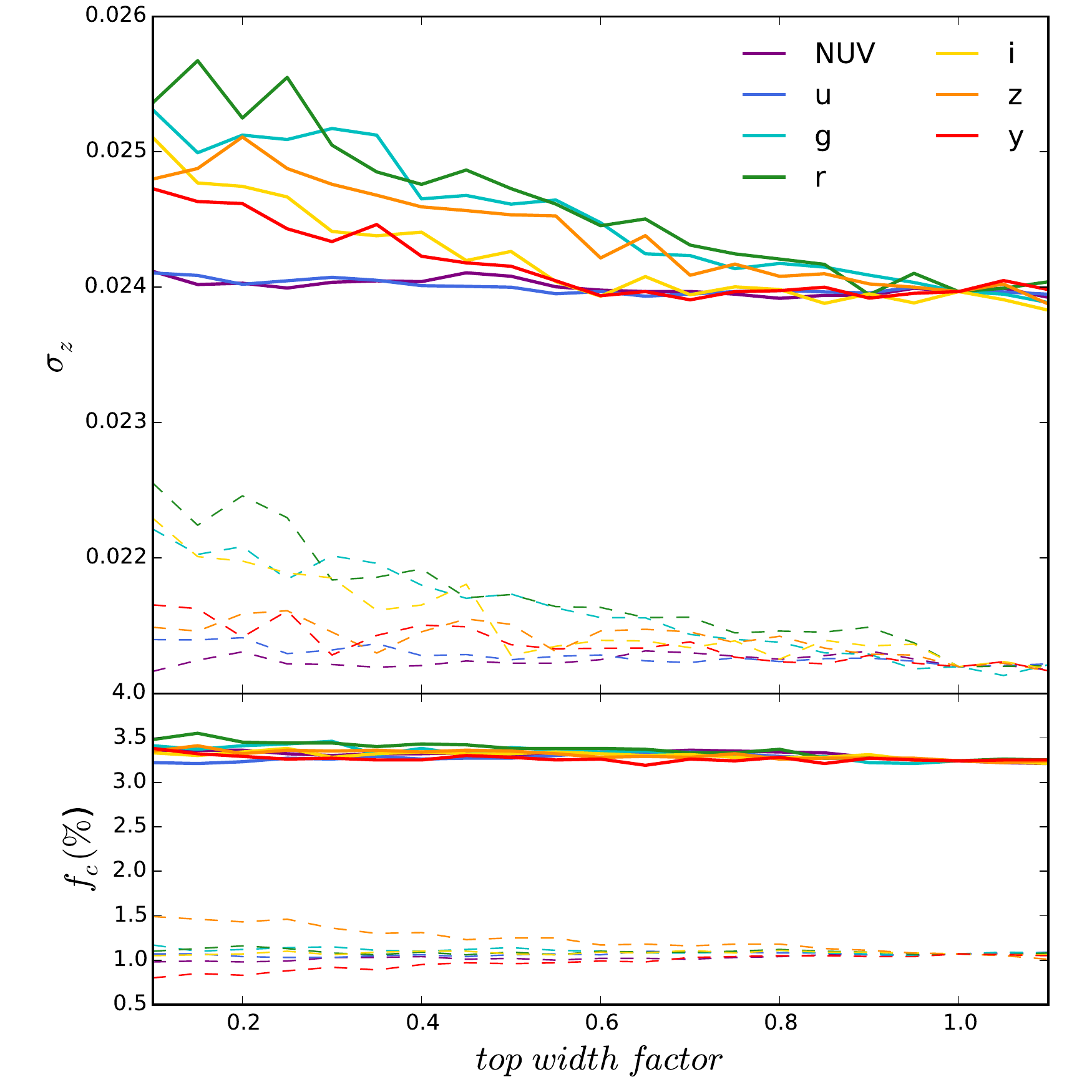}
\caption{The $\sigma_z$ and $f_{\rm c}$ as functions of the three filter parameters $\lambda_{\rm c}^x$, $\Delta \lambda^x$, and $s^x$. Here we show the results as functions of the shift scale of $\lambda_{\rm c}^x$ in the top panel, the width stretch scale $\Delta \lambda_{\rm s}^x$ in the middle panel, and the factor $b$ multiplied on the top width of transmission curve in the bottom panel. The solid and dashed curves are for the real and 100\% detector efficiencies, respectively. }
\label{fig:fp}
\end{figure}

Now we estimate the photo-$z$ accuracy obtained by the CSS-OS filters in certain filter parameter ranges, i.e. dependency of photo-z accuracy on filter parameters. This is also quite helpful for the filter design and manufacture. Here we focus on three main  parameters of filter transmission curves, i.e. the position of the central wavelength $\lambda_{\rm c}^x$ for band $x$, the band wavelength coverage or width $\Delta \lambda^x$, and the slope of transmission curve $s^x$. The strategy of changing these three parameters is shown in Figure~{\ref{fig:cha_stra}. This strategy can  efficiently avoid or reduce gaps and large overlaps between the changed bands.

In the top panel of Figure~\ref{fig:fp}, we show the $\sigma_z$ and $f_{\rm c}$ as a function of $\lambda_{\rm c}$. Here we use a shift scale in $\rm \AA$ to denote the positions of bands relative to the original positions given in Figure~\ref{fig:filters} and Table~\ref{tab:filters}, and the shift range is from $-400$ to $+400$ $\rm \AA$. We explore two cases of both the real and 100\% detector efficiencies. We find that the variation of the fitting results for the two cases are generally similar. The result for the original position case, where the shift scale is equal to $0$, is close to the minimum $\sigma_z$ and $f_{c}$. A bluer $NUV$ band or redder $u$ and $g$ bands may be better, but the improvements of $\sigma_z$ and $f_c$ are not much, about 0.001 and 1\% at most, respectively. It indicates that the original positions of the seven filters are proper for photo-$z$ calibration.

In the middle panel of Figure~\ref{fig:fp}, the $\sigma_z$ and $f_{\rm c}$ as a function of band width are shown. A width stretch scale $\Delta \lambda_{\rm s}$ is adopted to adjust the original band width $\Delta \lambda_{\rm ori}$, and the tested band width is given by $\Delta \lambda=\Delta \lambda_{\rm ori} + \Delta \lambda_{\rm s} $. Our original band widths, where the stretch scale $\Delta \lambda_{\rm s}=0$, are around the minimum $\sigma_z$ and $f_{c}$. We find that the $g$, $r$, and $i$ bands cannot be too wide, otherwise they will squeeze their adjacent bands to suppress $\sigma_z$. A wider $NUV$ band can be helpful to pin down $\sigma_z$ and $f_{c}$ by 0.001 and 1\%, respectively, at most, similarly as shifting the band position case.

In the bottom panel of Figure~\ref{fig:fp}, the results of varying the slope of the transmission curve are shown. To change the slope, we multiply a factor $b$ of the top width of the intrinsic transmission curve, while fixing the band FWHM. Then the top width $\Delta \lambda_{\rm top}=b\,\Delta \lambda_{\rm top}^{\rm ori}$, where $\Delta \lambda_{\rm top}^{\rm ori}$is the original width at 90\% of the maximum transmission. The $b$ range we explore is from 0.1 to 1.1, and the top width will be larger than the bottom width if $b\gtrsim1.1$ for all of the bands. We find that the original slopes for both real and 100\% detector efficiency cases are at the minimum values of $\sigma_z$ and $f_{\rm c}$. For shallower or steeper slopes where $0.9\lesssim b<1.1$, the changes of $\sigma_z$ and $f_{\rm c}$ can be neglected. For $b<0.5$, the variances of $\sigma_z$ and $f_{\rm c}$ become considerable.


\subsection{Comparison with other work}

\begin{figure}
\centering
\includegraphics[width=\columnwidth]{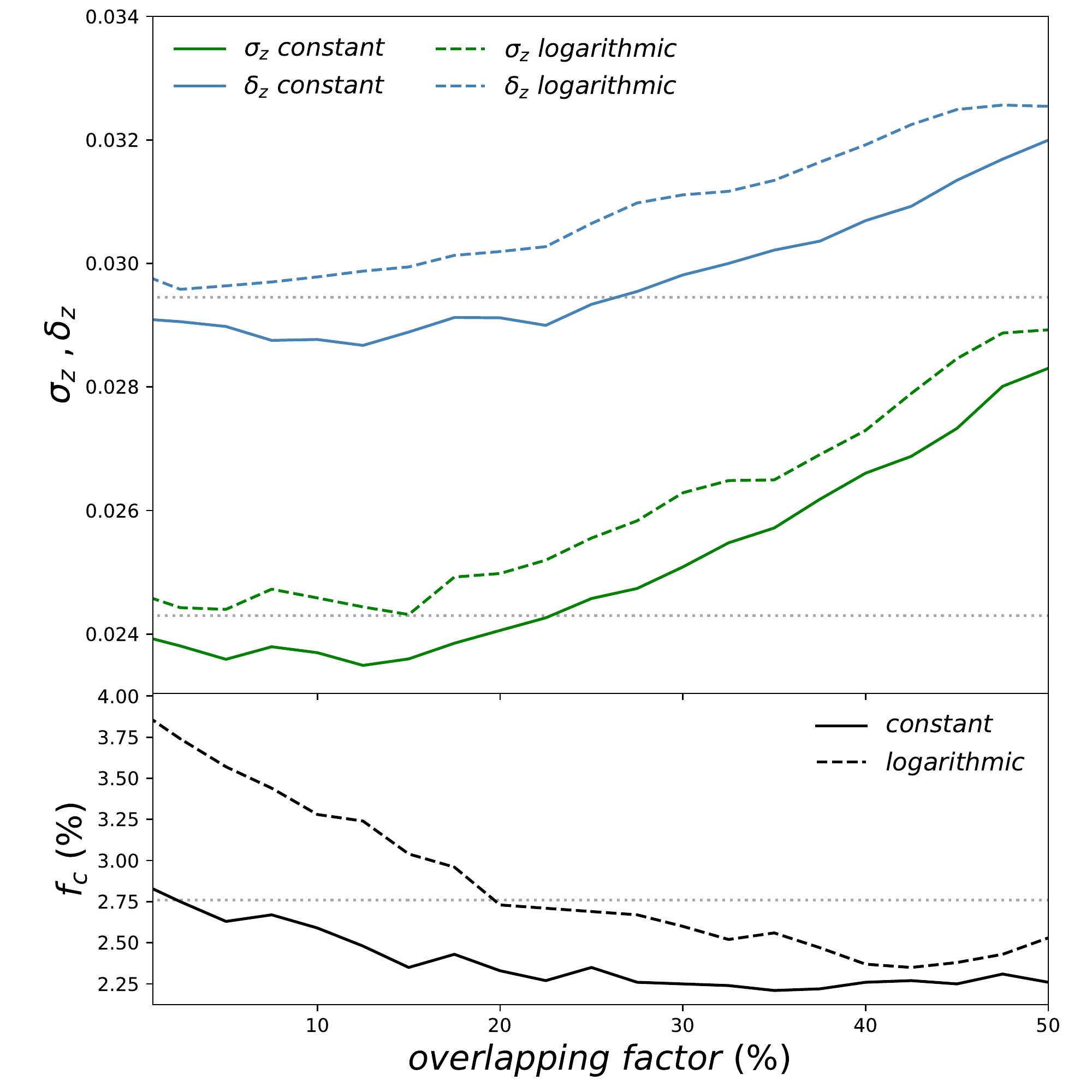}
\caption{The $\sigma_z$, $\delta z$, and $f_c$ as a function of overlapping fraction of the CSS-OS filters using real detector efficiency. The left and right ends of the curves show the results corresponding to the ``adjacent'' and ``overlapping'' cases in \citet{Benitez09}, respectively. The gray dotted lines denote the corresponding $\sigma_z$, $\delta z$, and $f_c$ derived from the current CSS-OS filter set.}
\label{fig:comparison} 
\end{figure}

In the last section, we find that the current CSS-OS filter can provide good $\sigma_z$ and $f_c$ in certain filter parameter ranges, that can meet the science requirement. In this section, we compare the CSS-OS filter sets to other types of sets (with different overlapping fractions) using the real CSS-OS detector efficiency (shown in the right panel of Figure~\ref{fig:filters}).

In \citet{Benitez09}, they explored how photo-$z$ performance depends on the number of filters $N_f$ with four types of filter sets, which are characterized by whether they have constant or logarithmically increasing band widths, and whether they have minimal or half-width overlaps\footnote{Note that they assumed constant SNR for the same AB magnitude by varying observing time per filter, which is different from the CSS-OS with fixed exposure time for each filter.}. They found that the filter systems with $N_f\gtrsim8$ perform much better than low $N_f$ systems in both completeness depth (defined by 80\% completeness magnitude limit $I_{80\%}$) and photo-$z$ accuracy. This demonstrates that the current CSS-OS filter set with $N_f=7$ can provide good photo-$z$ performance. Besides, after including the near-IR filters, i.e. $J$, $H$, and $K$ bands, they found that both the completeness magnitude and photo-$z$ accuracy can be further improved, which is consistent with our result given in subsection 4.1.

We also investigate the four types of filter sets in the CSS-OS frame with $N_f=7$ and wavelength coverage from 2500 $\rm \AA$ to 11000 $\rm \AA$ for 100\% detector efficiency case. Our results are shown in Figure~\ref{fig:comparison}. Here, instead of only minimal and half-band overlaps, we continuously increase the overlapping fraction of the filters with constant width or logarithmically increasing width to estimate the $\sigma_z$, $f_c$, and $\delta z$, which is the rms of $\Delta z/(1+z)$ excluding catastrophic ``outlier''. We find that both $\sigma_z$ and $\delta z$ have similar variation, which becomes bigger and bigger as the overlapping fraction increases, and the filter sets with constant widths have smaller $\sigma_z$ and $\delta z$ than the filters with logarithmic widths. These results are well consistent with that shown in Figure 4 of \citet{Benitez09} when $N_f\sim 7$. On the other hand, $f_c$ tends to become smaller when the overlapping fraction increases.

The corresponding $\sigma_z$, $\delta z$, and $f_c$ derived from the current CSS-OS filter definition are also shown in gray dotted horizontal lines, respectively. We find that the CSS-OS filter set, as expected, is similar with the logarithmic filter type with small overlaps (overlapping fraction less than 20\%). It is in a good agreement with the suggested filter set given by \citet{Benitez09} for $N_f<8$ (see their Figure 2 and 3). As can be seen in Figure~\ref{fig:comparison}, the current CSS-OS filter system can effectively suppress the fraction of catastrophic redshifts with $f_c\simeq2.76\%$, which is close to the smallest $f_c\simeq2.25\%$ given by constant-width filter sets with overlapping fraction between 20\% and 50\% (although the values of $\sigma_z$ and $\delta z$ are large in this range.). Besides, the current CSS-OS filter set also has sufficiently high photo-$z$ accuracy with $\sigma_z\simeq0.024$ and $\delta z\simeq0.029$, which almost reach the smallest $\sigma_z\simeq0.023$ and $\delta z\simeq0.029$ provided by the constant-width filter sets.

\section{Summary}

In this work, we test the photo-$z$ accuracy that can be measured by the CSS-OS in certain filter parameter ranges. The CSS-OS has seven filter passbands that cover a large wavelength range from the NUV to NIR bands. We adopt the COSMOS galaxy catalog with the similar magnitude limits as our survey. This catalog can represent our survey as real as possible, and it can provide similar observed galaxy redshift distribution, magnitude distribution, and galaxy types. Then we select galaxies with high quality based on the CSS-OS instrumental parameters, and use this selected sample in the photo-$z$ fitting process.

Then we calculate the mock observed flux and error for each band. We make use of thirty-one SED templates for elliptical, spiral, and young blue star forming galaxies. Since the CSS-OS has large wavelength coverage, we extend the wavelength coverages of the SED templates from $\sim900\ \rm \AA$ to $\sim90\ \rm \AA$ using the BC03 method. We also consider the dust extinction effect from both interstellar dust of galaxy and the absorptions of the IGM, when estimating the mock flux data. The flux error is evaluated by considering several factors, such as the instrumental parameters, sky background, and systematic errors. 

We compare three different photo-z SED template fitting codes: LePhare, EAZY, and Hyperz, and find that LePhare gives the best results for our survey. Furthermore, we improve the LePhare code by including the information of poorly detected data in the photo-$z$ fitting. We show that this can remarkably suppress the catastrophic redshift fraction and improve the photo-z accuracy. By applying this method, we find that  the CSS-OS photo-$z$ estimate can achieve $\sigma_z\sim0.02$ and $f_{\rm c}\sim3\%$.

Next, we explore the effect of photo-$z$ fitting accuracy for each band. By removing one band at a time, we perform the photo-$z$ fitting process, and calculate $\sigma_z$ and $f_{\rm c}$. We find that the $g$, $r$, and $i$ bands have the largest impact on both $\sigma_z$ and $f_{\rm c}$, and the $NUV$ band can affect $f_{\rm c}$ significantly. On the other hand, the $u$, $z$ and $y$ bands have relatively smaller effect on $\sigma_z$ and $f_{\rm c}$. Besides the seven bands in our survey, we also study other NIR bands at longer wavelengths for exploring the synergy with other surveys, i.e. the $J$ and $H$ bands.

Finally, we estimate photo-$z$ accuracy for the CSS-OS filter set within certain filter parameter ranges, and compare it with other types of filter sets. We find that the $\sigma_z$ and $f_c$ of the CSS-OS are always less than 0.03 and 5\% (around 0.02 and 3\%), respectively, in the filter parameter ranges we explore. Our results are also consistent with other work. Thus the current design can provide accurate photo-$z$ estimate, and could sufficiently meet the science requirement.

As expected, the CSS-OS will observe about one billion galaxies in the photometric imaging survey, and we find that about $58\%$ of them can have $\sigma_z\simeq 0.02$, and $\sim 95\%$ can reach $\sigma_z=0.05$ using the current filter definition based on our estimation. The photo-$z$ accuracy also can be further improved by including other information and methods, such as galaxy color and shape priors, and training sample method with spectroscopic data. Therefore, we believe that the current CSS-OS filter set can provide sufficiently accurate photo-$z$ measurement for the scientific goals, and could lead to exciting scientific discoveries in the future observation.

\section*{Acknowledgements}
\addcontentsline{toc}{section}{Acknowledgements}
YC and YG thank Yingjie Peng for helpful discussion. YG acknowledges the support of NSFC-11773031 and the Bairen program from the National Astronomical Observatories, Chinese Academy of Sciences. YG and XLC acknowledge the support of NSFC-11633004. CKX acknowledges the support of NSFC-11643003. XLC acknowledges the support of the MoST 863 program grant 2012AA121701, the CAS grant QYZDJ-SSW-SLH017, and the NSFC through grant No. 11373030. DZL acknowledges the support of NSFC-11333001 and 11173001. YQX acknowledges the support of NSFC-11473026 and 11421303. LC, XF, XZ, SW, and HZ were partially supported by the China Manned Space Program through its Space Application System and by the National key Research Program of China ``Scientific Big Data Management System'' (No.2016YFB1000605).

\appendix

\begin{figure*}
\centerline{
\resizebox{!}{!}{\includegraphics[scale=0.33]{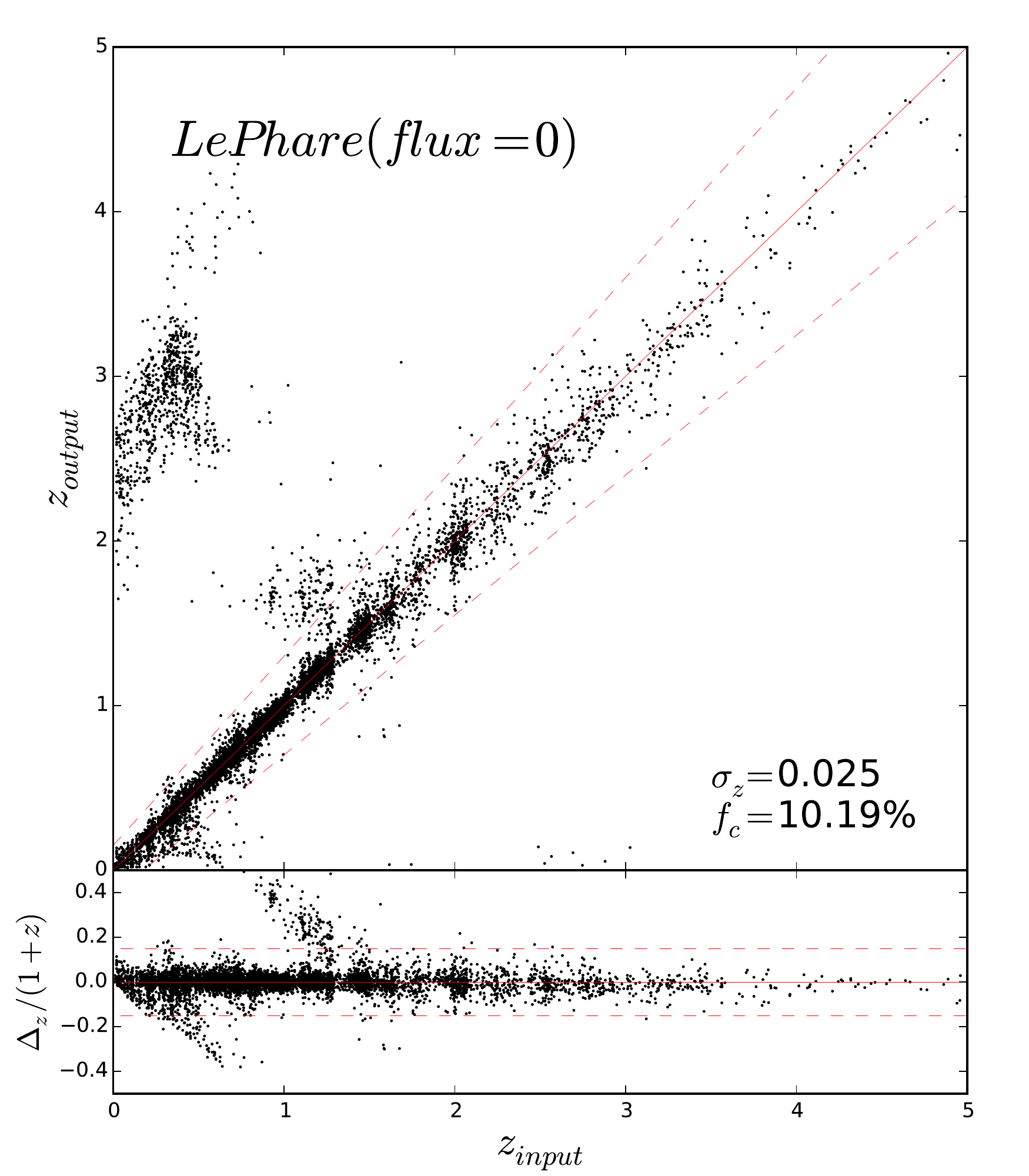}}
\resizebox{!}{!}{\includegraphics[scale=0.33]{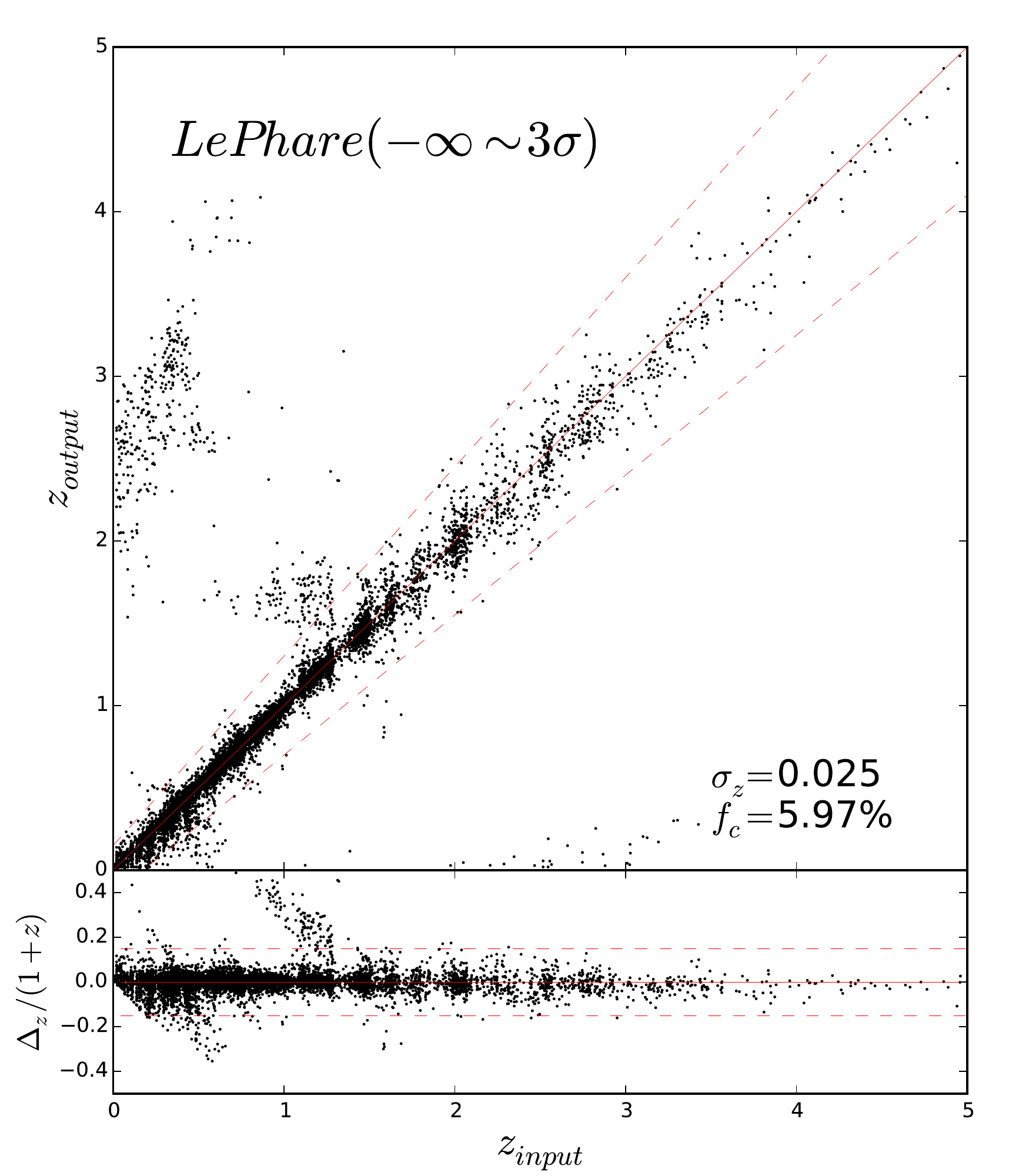}}
\resizebox{!}{!}{\includegraphics[scale=0.33]{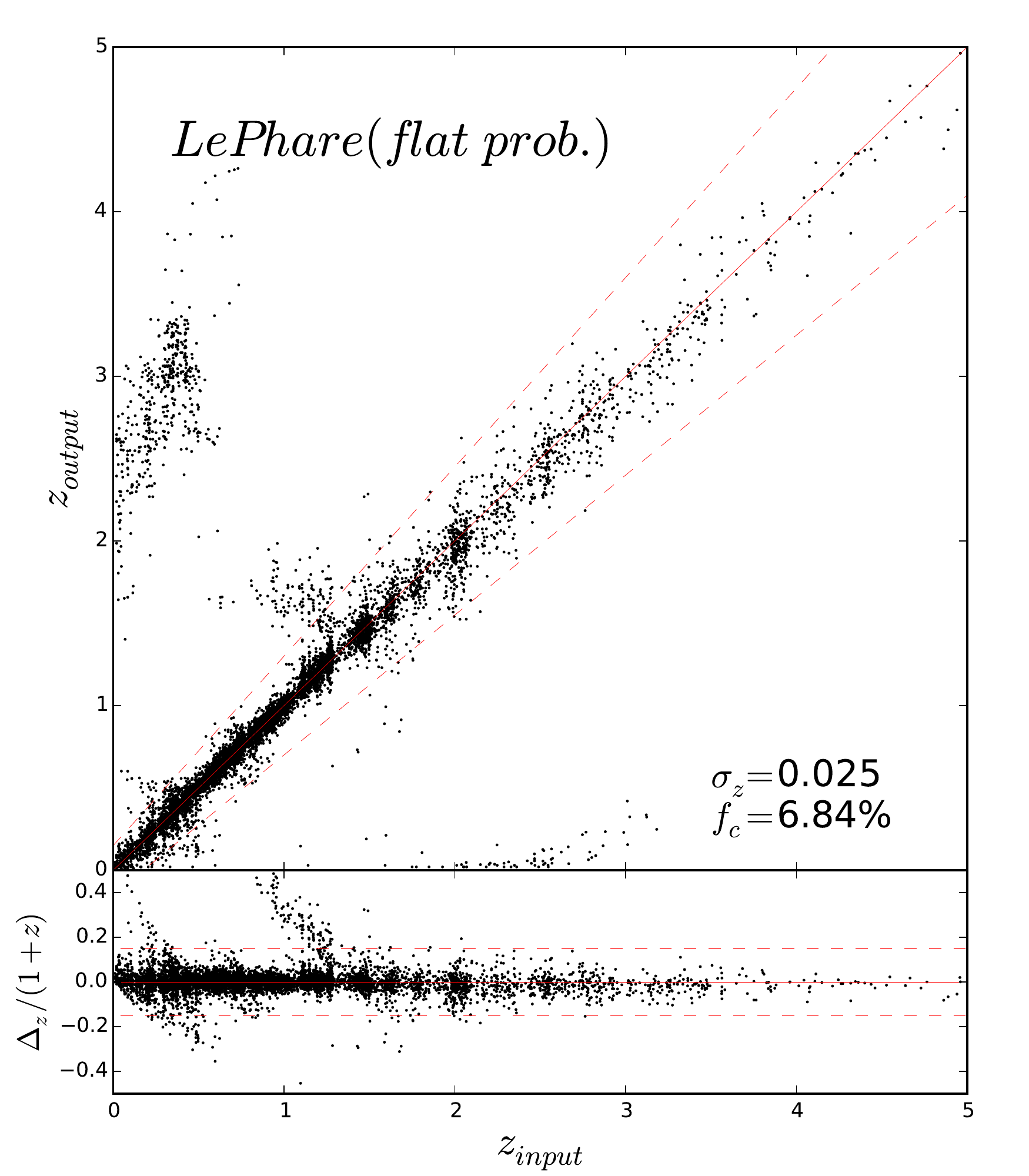}}
}
\caption{$Left:$ Result of setting flux and error to be 0  and $3\sigma$ magnitude limit, respectively, for the data with sensitivity below $3\sigma$ detection. $Middle$: Result from Eq.~(\ref{eq:chi2_tot}) and (\ref{eq:Pj}) but with $F_{\rm l}=-\infty$. $Right:$ Result of assuming flat accepting probability in $0\sim3\sigma$ flux range for poorly measured data below $3\sigma$ detection.}
\label{fig:pz_com}
\end{figure*}

\section{Comparison of different photo-$z$ fitting methods}

Besides the two photo-$z$ fitting methods shown in the main text, here we compare other three methods based on the Lephare code. 

The first one is setting flux and its error to be zero and $3\sigma$ magnitude limit, respectively, for the data with sensitivity below $3\sigma$ detection. As shown in the left panel in Figure~\ref{fig:pz_com}, we find that $\sigma_z=0.025$ and $f_c=10.19\%$, which has large catastrophic redshift fraction (although $\sigma_z$ is improved a little) that even larger than the result of removing the data in this band (see the left panel of Figure~\ref{fig:zp_code}). This is because that a bias can be introduced on the flux when enforcing it to be zero, since the measured flux in optical surveys probably has a positive value. Hence it will significantly affect the photo-$z$ fitting for the poorly measured data of the CSS-OS.

In the second method, we integrate from $F_{\rm l}=-\infty$ in Eq.~(\ref{eq:Pj}) instead of $F_{\rm l}=0$ shown in the main text. This method is usually used in infrared surveys, which always have large background. We find that $\sigma_z=0.025$ and $f_c=5.97\%$ as shown in the middle panel of Figure~\ref{fig:pz_com}. This $f_c$ is better than that of removing data method (left panel of Figure~\ref{fig:zp_code}), but worse than the results from the methods of integrating from $F_{\rm l}=0$ and using as-measured flux and error (see Figure~\ref{fig:LePhare_mod}). This is due to similar reason as the first method, that is introducing flux bias. Integrating from $F_{\rm l}=-\infty$ means that the largest probability of $P_j$ in Eq.~(\ref{eq:Pj}) is around flux=0, since the theoretical flux obtained from SED templates is always greater than 0. This is quite similar as the first method mentioned above, that a flux bias can be introduced in the photo-$z$ fitting process. But we can see that this effect is obviously reduced here, since a probability distribution is adopted in this method.

We also test the third method that treats $0\sim3\sigma$ flux range with flat probability instead of a Gaussian distribution used in Eq.~(\ref{eq:Pj}). As can be seen in the right panel of Figure~\ref{fig:pz_com}, we get $\sigma_z=0.025$ and $f_c=6.84\%$, which is a bit worse but comparable to the second method. This method can be performed by rejecting predicted flux if $F_j^{\rm th}>3\sigma$, for the data in band $j$ with sensitivity below $3\sigma$.

As the results of the six methods given in Figure~\ref{fig:pz_com}, \ref{fig:zp_code}, and \ref{fig:LePhare_mod}, we find that the two methods shown in Figure~\ref{fig:LePhare_mod} can provide the best photo-$z$ fitting results, and can be adopted in the CSS-OS data analysis.

\bsp	
\label{lastpage}
\end{document}